\documentclass[conference]{IEEEtran}
\IEEEoverridecommandlockouts
\usepackage{cite}
\usepackage{amsmath,amssymb,amsfonts}
\usepackage{algorithmic}
\usepackage{graphicx}
\usepackage{textcomp}
\usepackage{xcolor}
\usepackage{graphicx} 
\usepackage{epstopdf}
\usepackage[utf8]{inputenc}
\usepackage{array}
\usepackage{multirow}
\usepackage{graphicx}
\usepackage{subfigure} 
\def\BibTeX{{\rm B\kern-.05em{\sc i\kern-.025em b}\kern-.08em
		T\kern-.1667em\lower.7ex\hbox{E}\kern-.125emX}}
\begin{document}
	
	\title{When Machine Learning Meets Congestion Control:\\A Survey and Comparison\\
	}
	
   \author{Huiling~Jiang,~\IEEEmembership{Student   Member,~IEEE,}
	Qing~Li,~\IEEEmembership{Member,~IEEE,}
	Yong~Jiang,~\IEEEmembership{Member,~IEEE,}
	\\GengBiao~Shen,~\IEEEmembership{Student Member,~IEEE,}
	Richard~Sinnott,~\IEEEmembership{Member,~IEEE,}
	Chen~Tian,~\IEEEmembership{Member,~IEEE,}
	\\and~Mingwei~Xu,~\IEEEmembership{Member,~IEEE}
	\thanks{H. Jiang is with Tsinghua-Berkeley Shenzhen Institute, Tsinghua University, 518055 Shenzhen, China e-mail: jiang-hl19@mails.tsinghua.edu.cn}
	\thanks{Q. Li is with Institute of Future Networks, Southern University of Science and Technology, 518055 Shenzhen, China e-mail: liq8@sustech.edu.cn.}
	\thanks{Y. Jiang is with Computer Science and Technology, Tsinghua University, 100091 Beijing, China e-mail: jiangy@sz.tsinghua.edu.cn.}
	\thanks{G. Shen is with Computer Science and Technology, Tsinghua University, 100091 Beijing, China e-mail: gengbiao\_shen@126.com.}
	\thanks{R. Sinnott is with School of Computing and Information Systems, University of Melbourne, 3004 Melbourne, AUS e-mail: rsinnott@unimelb.edu.cn.}
	\thanks{C. Tian is with Computer science, Nanjing University, 210093 Nanjing, China e-mail: tianchen@nju.edu.cn.}
	\thanks{M. Xu is with Computer Science and Technology, Tsinghua University, 100091 Beijing, China e-mail: xumw@tsinghua.edu.cn.}}

	\maketitle
	
\begin{abstract}
		
Machine learning (ML) has seen a significant surge and uptake across many diverse applications. The high flexibility, adaptability and computing capabilities it provides extends traditional approaches used in multiple fields including network operation and management. Numerous surveys have explored ML in the context of networking, such as traffic engineering, performance optimization and network security. Many ML approaches focus on clustering, classification, regression and reinforcement learning (RL). The innovation of this research and contribution of this paper lies in the detailed summary and comparison of learning-based congestion control (CC) approaches. Compared with traditional CC algorithms which are typically rule-based, capabilities to learn from historical experience are highly desirable. From the literature, it is observed that RL is a crucial trend among learning-based CC algorithms. In this paper, we explore the performance of RL-based CC algorithms and present current problems with RL-based CC algorithms. We outline challenges and trends related to learning-based CC algorithms. 
\end{abstract}
	
\begin{IEEEkeywords}
	Congestion Control; Machine Learning; Reinforcement Learning; Learning-based
\end{IEEEkeywords}
	
\section{Introduction}
	
\IEEEPARstart{A}{s} a fundamental component of computer networks, CC plays a significant role in improving the network  resource utilization to achieve better performance. With the emergence of a large number of new technologies and new networks, e.g., data centers (DCs), WiFi, 5G and satellite communications, the complexity and diversity of network transmission scenarios and protocols have increased dramatically. This has brought  significant challenges to transmission protocol design. A rich variety of CC algorithms have been designed for specific scenarios. However, the variety of network scenarios and more importantly the intrinsic dynamics of the network, make it extremely difficult to design efficient generic CC algorithms. Therefore, CC algorithms based on ML have been proposed to provide a generic CC mechanism that could potentially underpin different network scenarios. In this paper, we provide a background analysis on traditional CC. Based on this, we investigate current works and research challenges in the application of ML in the field of CC.
	
\subsection{Traditional Congestion Control}

The Internet transmission protocol is based on packet switching over best-effort network forwarding \cite{137}. End-to-end transmission control is required to provide a reliable service for applications. To avoid network degradation caused by congestion, CC algorithms are typically employed to improve reliable transmission over the network. Network congestion occurs when excessive numbers of data packets are sent over the network by hosts \cite{138}. The objective of CC algorithms is to achieve  higher network throughput while avoiding packet loss caused by network overload. CC should ideally also guarantee the fairness between end-to-end sessions.

The traditional CC algorithms can be categorized into two types: end-to-end CC \cite{36.3,37.2,37.3} and network-assisted CC \cite{70,71,72}. End-to-end approaches only require the collaboration of senders and receivers, and hence they do not rely on any explicit signals from the network. Network-assisted approaches require support of  network devices, e.g. congestion information from routers. These are essential to achieve fairness and responsiveness in complex networking scenarios.

For end-to-end CC, one of the main challenges is to identify network congestion from implicit session signals, including packet loss and transmission delays. There are three main types of end-to-end CC approaches: loss-based CC, delay-based CC, and hybrid CC.

Generally, loss-based approaches such as \cite{36.5,36.6,36.7} adjust the sending rate when a given sender has not received a corresponding acknowledgement (ACK) over a given (long) time period, which typically indicates packet loss. Loss occurs when the buffer in a given network device is overloaded, thus loss-based approaches are supposed to attain high throughput through high bandwidth utilization. However, for some delay-sensitive applications, lower transmission times cannot be guaranteed. Besides, a packet loss may not be triggered by network congestion (e.g., random packet dropping), which may mislead any CC decision.

Therefore, delay-based approaches such as \cite{37.4,37.6,37.7} have been proposed. Delay-based approaches rely on detected transmission delays caused by the network. Compared with loss-based approaches, delay-based approaches are more suited for high-speed and flexible networks such as wireless networks, as they are not influenced by random packet loss. However, calculating the exact transmission delay remains a significant challenge. For example, a slight change in packet processing time in the host stack may cause deviations in the measured transmission delay, leading to erroneous decisions related to the sending rate.

To take full advantage of both loss and delay, hybrid approaches such as \cite{38.1, 38.3, 38.4} have been put forward. Although it is noted that these approaches  cannot identify the network status precisely based on implicit signals related to packet loss and transmission delay.

To solve this problem, network-assisted CC approaches such as \cite{134,136} have been proposed, where network devices provide explicit signals related to the network status for hosts to make sending rate decisions. When the network device is congested, some packets will be marked with a signal: Explicit Congestion Notification (ECN). The receiver will send back the ECN signal in the ACK and the sender will adjust the sending rate accordingly. The ECN signal for congestion is employed in \cite{135}. To further improve CC performance, multi-level ECN signals for congestion are employed in \cite{133}, which provides finer-grained CC.

With the emergence of a large number of new technologies and networks, e.g., DCs, WiFi, 5G and satellite communications, the complexity and diversity of network transmission scenarios have increased dramatically. This has brought significant challenges to CC. Whilst traditional CC approaches may work well in one scenario, they may not guarantee the performance in diverse network scenarios. Furthermore, the changing traffic patterns in one network scenario may also affect the performance of the solution. Therefore, an intelligent CC approach is required. 
	
\subsection{Learning-based Congestion Control}


The dynamic nature, diversity and complexity of network scenarios have brought significant challenges for CC. As such, it is difficult to design a generic scheme for all network scenarios. Furthermore, the dynamic nature of even the same network can make the performance of CC unstable. Current network environments may also include both wired networks and wireless networks, making the detection of packet loss more difficult \cite{83,84,86}. 

To solve the aforementioned problems, learning-based CC algorithms have been proposed. Different from traditional CC algorithms, learning-based schemes are based on real-time network states to make control decisions instead of using predetermined rules. This allows them to have better adaptability to dynamic and complex network scenarios.


Based on different mechanisms, learning-based CC algorithms can be divided into two groups: performance-oriented CC algorithms and data-driven CC algorithms. Performance-oriented CC algorithms employ objective-optimization methods to train the model and get the output. Generally, this kind of algorithms require manually effort to determine the parameters in utility function. The learning process is supposed to converge to the optimal value of the utility function. There are some typical performance-oriented CC algorithms. Remy \cite{23} is an early version among performance-oriented CC algorithms, whose utility function consists of throughput and delay. To maximize the expected value of the utility function, Remy finds the mapping based on pre-computed lookup table. Thence, corresponding sending rate is estimated. PCC \cite{12.3} and PCC Vivace \cite{145} show great performance as well based on designing utility functions which cover basic performance metrics such as the round-trip time (RTT). In \cite{146}, GCC applies Kalman filter which is a method that uses the linear system state equation to optimally estimate the system state through observation data. Based on Kalman filter, GCC estimates the end-to-end one way delay variation to dynamically control the sending rate. In \cite{147}, Copa optimizes the objective function based on current throughput and packet delay to adjust the sending rate. Compared with performance-oriented CC algorithms mentioned above, data-driven CC algorithms are more dependent on data sets and have problems with convergence. However, because data-driven CC algorithms update their parameters based on current data instead of relying on given constant parameters, they show stronger adaptability and satisfy diverse network scenarios through learning. Moreover, the mainstream research focuses more on data-driven CC algorithms. In this paper, our focus is on data-driven CC algorithms as well.

With regards to data-driven CC algorithms, machine learning techniques are used to train the model including supervised learning techniques, unsupervised learning techniques and RL techniques. Supervised and unsupervised learning techniques have been widely employed to improve network CC \cite{81, 83, 86}. However, these schemes are only partially successful because they are trained offline and are not capable of classifying realistic wireless and congestion loss \cite{84}. RL has more advantages in dealing with realistic congestion in networks with dynamic and sophisticated state space \cite{132, 12}. Therefore, RL techniques have been shown to be beneficial for CC because of the higher online learning capability \cite{7, 9}. At present, much research focuses on RL-based CC schemes.

However, learning-based CC is still in its infancy. Most learning-based CC algorithms adjust the congestion window (CWND) to control the sending rate instead of adjusting the sending rate directly. Therefore, burstiness is still a problem in high speed networks because the CWND can increase sharply when multiple ACKs arrive \cite{68}. Current learning-based CC algorithms such as \cite{17, 19} generally focus on end-to-end CC instead of network-assisted CC. Designing a general purpose learning-based CC scheme that can work in real network scenarios is still a major goal of both academia and industry.
	
\subsection{Overall Analysis}

In addition to considering current learning-based CC algorithms and providing systematic analysis and comparison, we conduct comprehensive experiments of learning-based CC under diverse dynamic network scenarios and compare them with more traditional algorithms. The implementation of learning-based CC algorithms in real network stacks has shown that they are often lacking because intelligent learning decisions cannot be made fast enough, i.e. in the order of 100 milliseconds with a GPU with 1Gb real network data transmission. Therefore, in order to judge the pros and cons of decision models we conduct comprehensive experiments of various schemes using the NS3 emulator \cite{128}. 

In the simulation, we compare the RL-based CC algorithms of Deep Q Learning (DQL)  \cite{52}, Proximal Policy Optimization (PPO) \cite{5} and Deep Deterministic Policy Gradient (DDPG) \cite{131} with the traditional CC algorithm  NewReno \cite{141}.
We design three different scenarios with different configurations of bandwidth and delay. The network with high bandwidth and low delay simulates a typical data center networks. The network with low bandwidth and high delay simulates typical wide area networks. The network with low bandwidth and low delay simulates  ad hoc wireless networks. These three network environments represent the diverse environments needed for learning-based CC algorithms.
In order to fully evaluate the performance of learning-based CC schemes, we generate network traffic traces with 80\% elephant flows and 20\% mice flows for experiments. The experimental results show that learning-based CC algorithms are more suitable for dynamic environments with higher Bandwidth Delay Product (BDP). For networks with low BDP, i.e. the link bandwidth is low or the link delay is low, learning-based CC algorithms are too aggressive to learn and deal with dynamic network stability. Moreover, the performances of these three learning-based CC algorithms shows no difference in our simulated environments because the complexity of the environments are limited. Therefore, all of them can handle these network scenarios.

In realistic scenarios, RL-based CC algorithms are influenced by the computation time needed for RL. This impacts the feasibility of these schemes. Therefore, we propose three potential solutions to deal with this problem. Firstly, designing lightweight models based on mapping tables of states and actions to decrease the time consumption of learning decisions. Secondly, decreasing the frequency of decisions to provide better feasibility under low-dynamic network scenarios. Finally, asynchronous RL can improve the decision speed of RL-based CC algorithms. 

Based on this analysis, we further explore the challenges and trends for future works in the area of learning-based CC. Current challenges of learning-based CC algorithms are mainly focused on engineering related issues such as parameter selection, high computational complexity, high memory consumption, low training efficiency, hard convergence and incompatibility. In the future, learning-based CC needs to receive more attention both from academia and industry. Based on the understanding and analysis of the current learning-based CC solutions, we identify trends in learning-based CC. First, because of their capability for   dealing  with network congestion with  dynamic  and sophisticated  state  spaces, RL-based CC will be a significant research trend moving forward. Second, given the excessive time and cost of learning decisions, lightweight learning-based CC will be a key research direction. Third, an open network test platform that provides massively differentiated dynamic network scenarios to support the exploration and verification of learning-based CC mechanism,  requires further contributions in the study of learning-based CC algorithms.

The rest of the paper is structured as follows. In Section II, we present related background knowledge. In Section III, IV and V, we consider supervised learning-based CC algorithms, unsupervised learning-based CC algorithms and RL-based CC algorithms respectively as representatives of three main groups of learning-based CC algorithms. In Section VI, we provide an overview of the setup of simulations. In Section VII, we conduct simulations and compare performances between RL-based CC algorithms and traditional CC algorithms. In Section VIII, we outline challenges and trends  of learning-based TCP. Finally in Section IX, we conclude the paper.
	
\section{Background}  
	
\subsection{CC mechanisms}
	
CC mechanisms typically involve four key issues: slow start, congestion avoidance, re-transmission and fast recovery \cite{139}. To illustrate the procedure of CC, we adopt the window-based CC. The sliding CWND determines the next packet to be sent.
	
\textbf{Slow Start.} At the initial stage of transmission, due to the unknown network transmission capability, CWND starts with a low value to prevent congestion caused by a large amount of data being injected to the network in a short period of time. This process is called slow start. In the classic slow start process, if an ACK is not delayed, each time a good ACK is received, it means that the sender can send twice the numbers of packets last sent, which will cause the sender's window to grow exponentially over time. Normally, a link buffer is under-loaded because the in-flight data is limited. Therefore, slow start can improve the link utilization due to the increasing speed.
	
\textbf{Congestion Avoidance.} In the slow start phase, CWND can grow rapidly, to a given threshold. Once the threshold is reached, it means that there may be more available transmission resources. If all resources are occupied immediately, severe packet loss and re-transmissions will occur on other connections sharing the queue of the router, resulting in unstable network performance. In order to get more transmission resources without affecting the transmission of other connections, TCP implements a congestion avoidance strategy. Once the slow start threshold is established, TCP will enter the congestion avoidance phase, and increase the value of CWND each time based approximately on the size of the successfully transmitted data segment. The increasing speed is much slower than the slow-start exponential growth. More precisely, CWND will update as follows for each new ACK:
	
\begin{equation}
CWND_{t+1}=CWND_t+SMSS*SMSS/CWND_t
\end{equation}
	
\textit{SMSS} is the maximum segment packet size of the sender. With the arrival of each ACK, CWND will have a small increase, and the overall growth rate will be slightly sub-linear. This growth process has been termed additive increase. Through this process, if congestion is detected, CWND will be reduced by half. 
	
\textbf{Re-transmission.} Re-transmission includes timeout re-transmission and fast re-transmission. Timeout re-transmission starts a timer after sending a given packet. If no acknowledged packet of the datagram is sent within a certain period of time, the data is re-transmitted until the transmission is successful. A key parameter that affects the efficiency of the timeout re-transmission protocol is the re-transmission timeout (RTO). Setting the value of RTO too large or too small will adversely affect the protocol. Fast re-transmission requires the receiver to send a duplicate ACK immediately after receiving an out-of-sequence segment so that the sender knows as soon as possible that there is a segment that has not reached the designated server, rather than waiting to send data confirmation itself. The re-transmission mechanism in CC ensures that data can be transmitted from the sender to the receiver.
	
\textbf{Fast recovery.} Fast recovery means that when the sender receives three duplicate ACKs in succession, it executes a multiplication reduction algorithm and halves the slow start threshold to prevent network congestion. The CWND increases slowly and linearly. The CWND then increases in an accumulative manner, causing the CWND to increase slowly and linearly. The fast recovery algorithm can avoid congestion and gradually reduce the window to affect the link utilization. 

Among traditional CC algorithms, the above four mechanisms make up the basic approaches while learning-based CC algorithms do not adopt strict rules to control congestion. To guarantee flexibility for different scenarios, learning-based CC algorithms can however learn different strategies to adjust the CWND instead of following fixed rules. 

\subsection{Rate Adjustment Mechanisms of Congestion Control algorithms}

To control the sending rate of input data, there are three rate adjustment mechanisms in CC algorithms: window-based techniques, rate-based techniques and pacing. 

Window-based strategies adjust CWND directly. CWND reflects the transmission capacity of the network. The actual window of the sender is the smaller of the CWND and the window of the receiver. Considering the convenience of window-based strategies, there are multiple traditional CC algorithms such as the classic algorithm DCTCP \cite{68}. Though window-based techniques are efficient, burstiness is a big issue especially in networks with high bandwidth. When a bunch of ACKs arrive, CWND will increase dramatically. Thus window-based strategies can result in variations, low throughput and high delay. 

Rate-based strategies control the actual sending rate directly, so they are able to fully make use of the bandwidth without burstiness. There are many  rate-based strategies. In \cite{62}, an early version of a rate-based strategy was presented to control congestion in asynchronous transfer mode (ATM) services.  \cite{63} combined control theory with rate-based strategies   to deal with flow control in continuous-time networks. However, because rate-based strategies rely on pre-designed rules that can adjust in each interval, the responsiveness is relatively lower compared with window-based strategies. Moreover, the complex rate-based strategies are often resource-consuming. 

Therefore, a hybrid strategy was presented based on packet pacing in \cite{65}. Packet pacing is acknowledgement-driven, which is similar to window-based strategies. As a result, responsiveness is guaranteed. In addition, based on packet pacing strategies, senders can allocate transmission tasks in given time intervals and hence  burstiness can be avoided. In \cite{67}, packet pacing strategies were shown to avoid the burstiness caused by bunches of arriving ACKs. However, packet pacing performs worse in throughput and fairness in some network scenarios including the initial period of the TCP communication \cite{61}. 

As shown above, different adjustment strategies can satisfy diversified network scenarios. Among traditional CC algorithms, most algorithms are window-based. With the development of CC algorithms, more and more rate-based CC algorithms and pacing techniques are designed. Based on the literature, most learning-based CC algorithms adopt window-based CC algorithms.

\subsection{Performance Metrics of Congestion Control algorithms}
	
CC algorithms are expected to achieve various goals and objectives as shown in Table I.
	
	\begin{table*}[h]
		\caption{Objectives of learning-based CC algorithms}
		\centering
		\renewcommand{\arraystretch}{1.5}
		\begin{tabular}{c|l}
			\hline
			Objective                              & \multicolumn{1}{c}{Description}                        \\ \hline
			Maximizing throughput                  & \begin{tabular}[c]{@{}l@{}}To maximize throughput, bandwidth utilization is supposed to be high. High throughput contradicts low \\ RTT or flow completion time since high throughput means the environment tolerates high queue lengths, \\ which may cause long delays.\end{tabular}                                         \\ \hline
			Minimizing RTT or flow completion time & \begin{tabular}[c]{@{}l@{}}Minimizing RTT or flow completion time is a basic requirement expected to be met. For each task, the flow \\ completion time reflects the delay, which is supposed to be as small as possible.\end{tabular}                                                                                              \\ \hline
			Minimizing packet loss rate            & \begin{tabular}[c]{@{}l@{}}Minimizing the packet loss rate is a basic goal of CC algorithms. Low packet loss rate means that there. is \\ a stable network environment and low delay.\end{tabular}                                                                                                                                                                                                  \\ \hline
			Fairness                               & \begin{tabular}[c]{@{}l@{}}Fairness is important for  multiple users. Resource allocation should be as fair as possible between users \\ and consider diverse applications.\end{tabular}                                                                                                                            \\ \hline
			Responsiveness                         & \begin{tabular}[c]{@{}l@{}}Updating the frequency and  adjustment policy of CWND can influence the responsiveness of algorithms. High \\ responsiveness is expected, which implies high resource-consumption as well. Therefore,  \\ responsiveness needs to be balanced based on different scenarios.\end{tabular} \\ \hline
		\end{tabular}
	\end{table*}
	
Throughput represents the amount of data that passes through a network (or channel, interface) in a given time interval. High throughput means high link utilization. Maximizing throughput is crucial. Given the link bandwidth, high throughput indicates high efficiency in transferring data. 
	
RTT measures the time including the transmission time, the propagation time, the queue time and the processing time. Flow completion time (FCT) indicates the time required to transfer the flows. RTT and FCT are expected to be small. For users, RTT and FCT show the delays that they may have to tolerate. However, it may be the case that maximizing throughput and minimizing RTT or FCT can be orthogonal. High throughput means making use of the link bandwidth as much as possible, which can give rise to an increased  queue length that may cause delays. 
	
The packet loss rate indicates the efficiency of the data transmission. For CC, minimizing the packet loss rate is important as it shows the control capability and stability of the network. 
	
Fairness is a measure of equality of the resource allocation of the network. Increased fairness requires CC algorithms to fairly allocate resources between flows to user's satisfaction and in turn improve the Quality of Service (QoS).

Responsiveness reflects the speed of the CC to deal with real-time flows. A high responsiveness level means that the algorithms can detect the congestion quickly and rapidly adjust the CWND to an optimal value.
	
These objectives are important for all CC algorithms, but they are hard to achieve. To get good performance for some targets, can mean that others have to  be sacrificed. In different scenarios, the targets may also have different priorities and hence trade-offs are necessary. Based on the previous literature, different CC research focus on different performance aspects including: throughput, RTT and the packet loss rate. In our simulations, we measure these three parameters in detail.
	
\section{Supervised Learning-based Congestion Control Algorithms}
	
In this section, we introduce supervised learning-based CC algorithms. Supervised learning techniques train given samples to obtain an optimal model, and then use this model to map all inputs to corresponding outputs. By performing judgments on the outputs and their ability to achieve  classification, supervised learning techniques have the ability to perform data classification. Classic supervised learning methods include decision trees, random forests, Bayes, regression and neural networks. 

In the networking domain, supervised learning methods are used to predict congestion signals for end-to-end networks and manage queue length for network-assisted networks. Congestion signal prediction consists of loss classification and delay prediction.  As mentioned before, congestion is detected implicitly based on packet loss or delay when congestion occurs  in traditional CC algorithms. In supervised learning-based CC algorithms, congestion is estimated in advance based on current and previous network states such as the queue length and the network delay. The key basis for this approach is that network states form a continuous time series, where the future state can be predicted by past states. Through this, supervised learning-based CC algorithms can be more intelligent compared with traditional CC algorithms. 
	
\subsection{Congestion Detection in End-to-end Networks}  
	
\subsubsection{Loss Classification}
	
Loss is a crucial but indirect signal used to detect congestion. It  gives nodes feedback in networks only when  congestion has already happened. In addition, basic loss-based CC algorithms cannot distinguish the cause of packet loss. Therefore, the classification of loss is essential to understand CC. 
	
Wireless networks provide many classic scenarios required to distinguish the wireless loss and congestion loss. In wireless networks, loss may be caused by erroneous wireless links, user mobility, channel conditions and interference. There has been a body of research related to loss classifications in wireless networks based on traditional CC algorithms. In \cite{115}, the proposed algorithm (Biaz) use the packet inter-arrival time to classify wireless loss and congestion loss. If the packet inter-arrival time is confined to a range, the missing packets are lost due to wireless loss. Otherwise, the loss is considered congestion loss. In \cite{116}, a new designed loss classifier for relative one-way trip time (ROTT) was used (Spike) to differentiate loss types. If the connection of ROTT was relatively higher, the loss was supposed to be caused by congestion. In other cases, the loss was assumed to be wireless loss. In \cite{77}, the amount of losses and ROTT were used to distinguish the types of loss. The presented algorithm, provided a hybrid algorithm (ZigZag) that was more efficient than the above two algorithms.
	
These loss classifiers are effective in some specific scenarios but have their limitations. Biaz \cite{115} is suitable for wireless last hop topology instead of the wireless bottleneck links with competitive flows while Spike \cite{116} shows better performance in wireless backbone topology with multiple flows. ZigZag \cite{77} is relatively more general, and hence is able to satisfy different topology scenarios however it is sensitive to the sending rate.

Considering the limitations of traditional loss classifiers for wireless networks, supervised learning techniques offer several advantages. To fully understand the loss information, multiple parameters can be taken into consideration. In \cite{83}, the one-way delay and inter-packet times were used as states to predict  loss categories. In \cite{84}, the queuing delay, the inter-arrival time and lists of packets were used as inputs. In addition, diverse supervised learning techniques were applied. In \cite{85}, decision trees, decision tree ensembles, bagging, random forests, extra-trees, boosting and  multi-layer perceptrons were used to classify the types of loss. Simulations show that these intelligent loss classifiers achieve high accuracy in different network scenarios.
	
Beyond wireless loss, contention loss is common in Optical Burst Switching (OBS) networks. OBS provide an advanced network, which saves the sources due to wavelength reservation. However, because of the lack of buffers in OBS, contention loss is generated when there is a burst at the core nodes. There are some supervised learning-based CC algorithms designed to tackle this. In \cite{114}, some classic contention resolutions are discussed and measured including wavelength conversion, deflection routing selection and buffering with shared feedback fiber delay line. To measure the efficiency of these strategies, burst loss probability and burst probability were considered. These strategies show good performances related to OBS contention issues. While in \cite{87}, a Hidden Markov Model was used to classify contention loss, congestion loss and control congestion separately. Simulations showed the effectiveness of loss classifiers in different network scenarios.
	
Reordering loss cannot be ignored in networks with multi-channel paths. In networks, when packets are reordered, reordering loss occurs. Supervised learning-based CC algorithms are able to deal with the associated classification issues. In \cite{76}, out-of-order delivery causes  variations of RTT. Therefore, RTT related with reordering and RTT related with congestion show different distributions. In \cite{86}, a Bayesian algorithm was used to represent the distributions of RTT for two types of losses. The proposed algorithm showed high prediction accuracy.  
	
In conclusion, wireless loss, contention loss and  reordering loss impact the detection of congestion loss. Supervised learning techniques show advantages in classifying types of losses in different network scenarios. The mechanism is shown in Figure 1 and Table II summarizes the studies related with loss classifiers based on supervised learning methods. However, there are some issues related with these supervised learning-based CC algorithms. 
	
Mis-classification is one issue. In wireless networks, predefined parameters determine the errors in classifying  congestion loss and  wireless loss. If the congestion loss is more easily classified than wireless loss, the classifier shows bad performances in wireless networks since the network is supposed to react when loss is detected. However, due to the mis-classification, the network considers congestion loss as wireless loss and does not control the sending rate quickly. Therefore,  congestion can not be reduced. Otherwise, if the wireless loss is more easily classified as  congestion loss, the algorithm is ineffective in wireless scenarios because there exists considerable wireless losses. As a result, the wireless network may overreact to loss signals. Therefore, parameters in the algorithms need to be considered carefully to balance  performance in different network scenarios.
	
The balance between computational complexity and prediction accuracy is another issue. As shown in \cite{85}, compared with decision trees,  boosting algorithms achieve higher accuracy but consume much more network resources. Therefore, considering the limited improvements in accuracy of boosting,  decision trees show more advantages, although there is always a trade-off.  

\begin{figure}[h]
	\centering  
	\includegraphics[height=4.7cm,width=9cm]{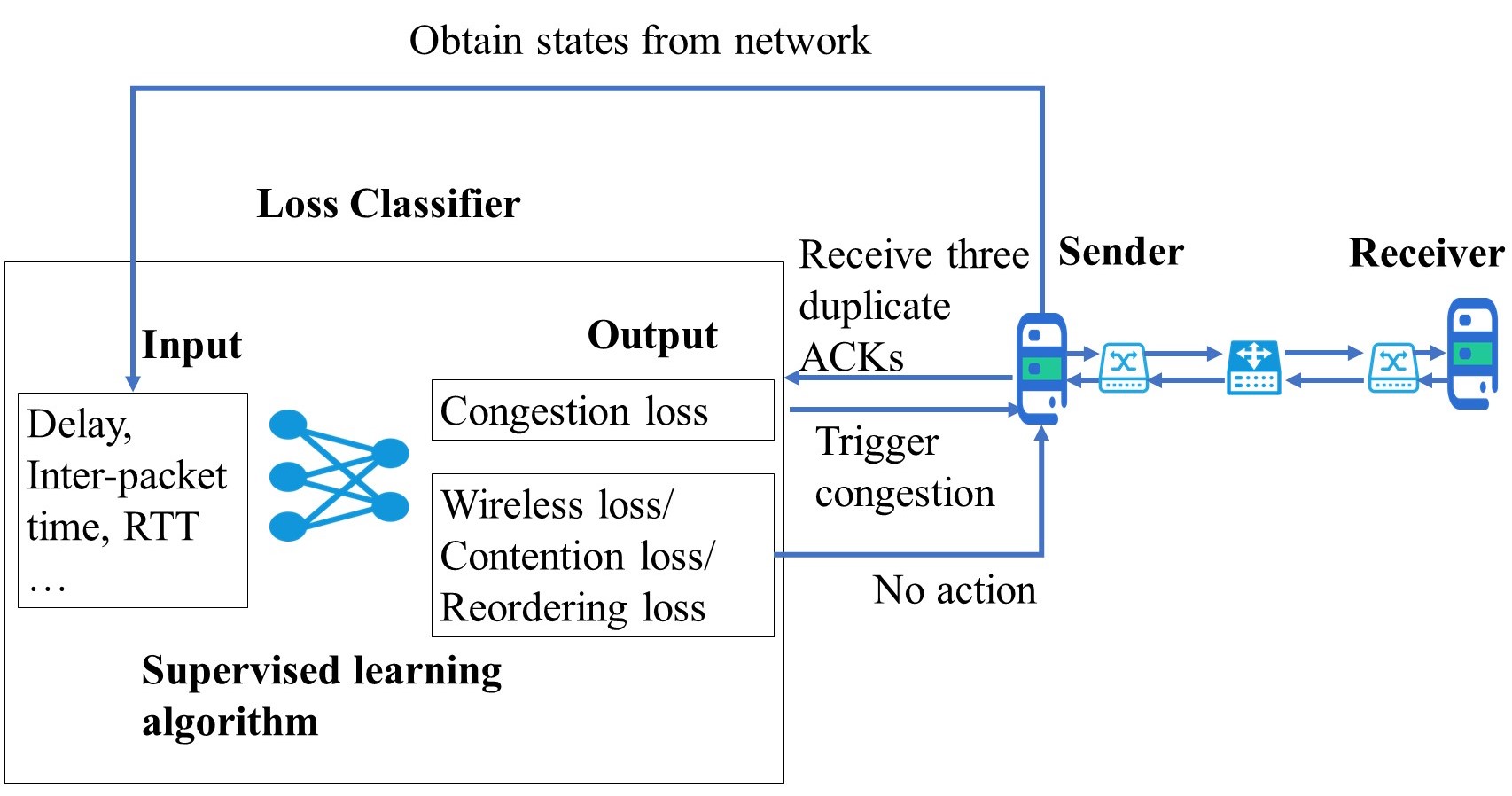}
	\caption{\textbf{Loss Classification Based on Supervised Learning Algorithms}}  
	\label{1}  
\end{figure}

	\begin{table*}[h]
		\centering
		\caption{Supervised Learning: Loss Classification in End-to-end CC Algorithms}
		\renewcommand{\arraystretch}{1.5}
		\begin{tabular}{c|c|l|c}
			\hline
			Algorithms                                                                                                                                                                                                                    & Scenarios                      & \multicolumn{1}{c|}{Input}                                                                                                                                 & Output                             \\ \hline
			Decision Tree Boosting \cite{83}                                                                                                                                                                             & Wireless networks              & One-way delay, inter-packet times                                                                                                                     & Link loss or Congestion loss       \\ \hline
			Bayesian \cite{86}                                                                                                                                                                                           & Networks with Reordered events & RTT of lost packets                                                                                                                                        & Reordering loss or Congestion loss \\ \hline
			Hidden Markov Model \cite{87}                                                                                                                                                                                & Optica Burst Switching         & \begin{tabular}[c]{@{}l@{}}The number of bursts successfully received \\ at an egress between any two bursts\end{tabular}                                  & Contention loss or congestion loss \\ \hline
			\begin{tabular}[c]{@{}c@{}}DT, Bagging, Boosting, \\ Neural Networks \cite{84}\end{tabular}                                                                                                                  & Wireless networks              & Queuing delay, inter-arrival times, lists of packets                                                                                                       & Wireless loss or Congestion loss   \\ \hline
			\begin{tabular}[c]{@{}c@{}}Decision Trees, \\ Decision Tree Ensembles, \\ Bagging, \\ Random Forests, \\ Extra-trees, \\ Boosting, Multilayer \\ Perceptrons,  \\ K-Nearest neighbors \cite{85}\end{tabular} & Wireless networks              & \begin{tabular}[c]{@{}l@{}}The standard deviation, the minimum, and the \\ maximum of the one-way delay, inter-packet time \\ for the packets\end{tabular} & Wireless loss or Congestion loss   \\ \hline
		\end{tabular}
	\end{table*}
	
\subsubsection{Delay Prediction}
	
As a congestion signal, the delay of transmissions reflects the amount of in-flight data, which shows the overall load on the network. There are some classic delay-based CC algorithms such as Vegas that measures delay accurately \cite{37.2}. However, in dynamic networks, traditional delay-based CC algorithms are not flexible enough. As Figure 2  shows  and Table III  concludes, supervised learning techniques have high learning capabilities and are efficient in predicting future delays and reacting quickly to avoid congestion.
	
RTT prediction is a major topic in delay prediction. Based on the measured RTT, other parameters can be calculated such as RTO. There has been a body of research exploring the prediction of RTO based on RTT. In \cite{108}, estimation of RTT was dynamically changed to estimate RTO in wireless network. In \cite{107}, RTT was used to predict RTO and bandwidth utilization. In \cite{105}, a fixed-share expert was used to compute the RTO in mobile and wired scenarios relying on RTT estimations. In addition, in \cite{110} and \cite{113}, the fixed-share leveraged exponentially weighted moving average technique demonstrates a more accurate algorithm.
	
Moreover, there has been various research measuring RTT based on other parameters in the network. In \cite{36}, linear regression was used to establish the relationship between RTT and the sending rate. In \cite{53}, a Bayesian technique was used to simulate the distribution between delay and the sending rate and then to predict delay based on the sending rate. This is needed in real-time video applications and wireless networks.

Delay prediction is also significant for delay-sensitive networks that require  networks with increased responsiveness. Several intelligent algorithms for the prediction of RTT using limited parameters and simple techniques to guarantee the low computational complexity and high responsiveness have been proposed. Further research is needed to push the boundary and deal with more complex related parameters and techniques to improve delay predictions.

\begin{figure}[h]
	\centering  
	\includegraphics[height=4.7cm,width=9cm]{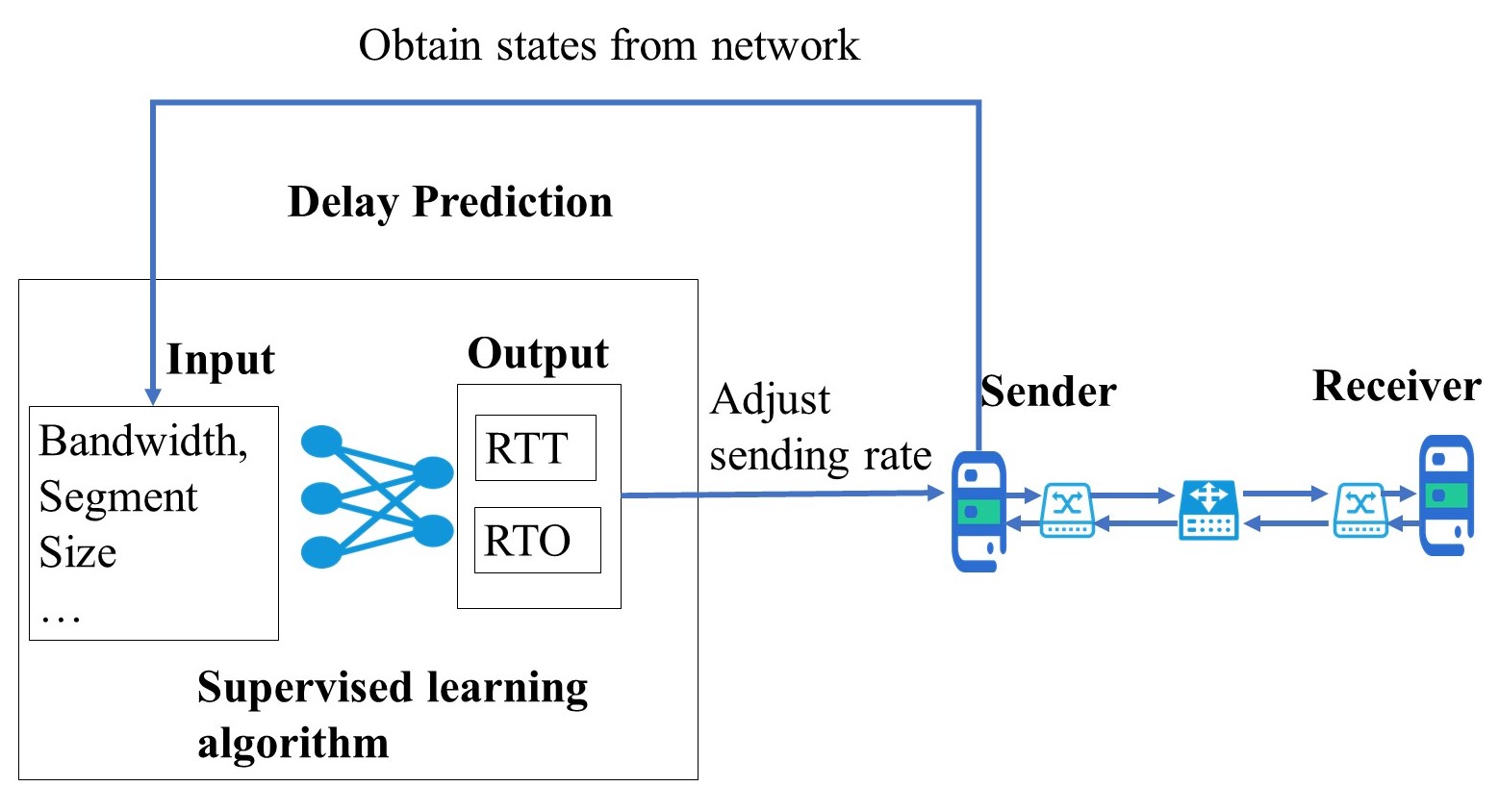}
	\caption{\textbf{Delay Prediction Based on Supervised Learning Algorithms}}  
	\label{2}  
\end{figure} 

\begin{figure}[h]
	\centering  
	\includegraphics[height=4.7cm,width=9cm]{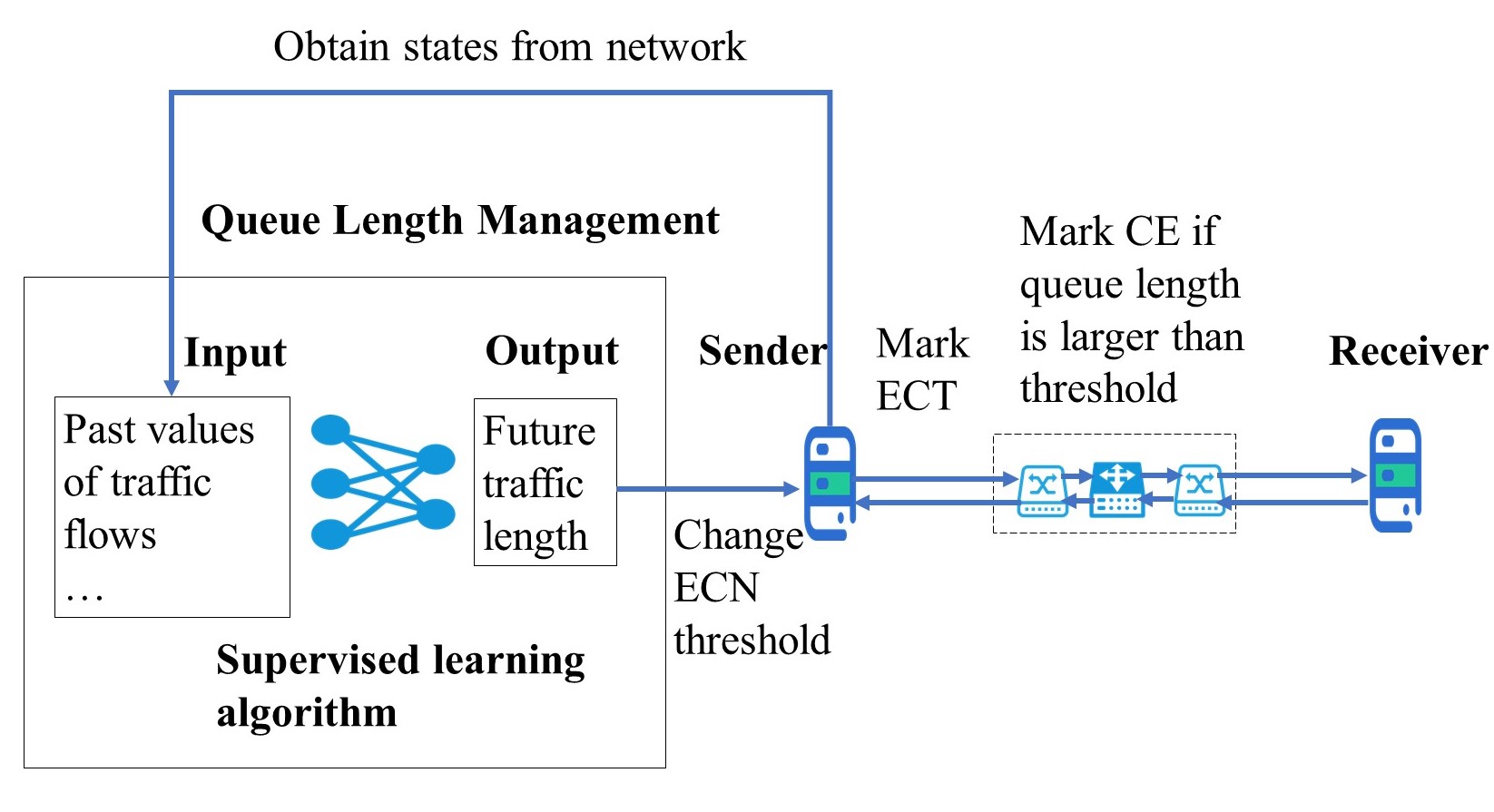}
	\caption{\textbf{Queue Length Management Based on Supervised Learning Algorithms}}  
	\label{3}  
\end{figure} 
	
	\begin{table*}[h]
		\centering
		\caption{Supervised Learning: Delay Measurement in End-to-end CC algorithms}
		\renewcommand{\arraystretch}{1.5}
		\begin{tabular}{c|c|l}
			\hline
			Algorithms                                              & Scenarios                                                                                     & \multicolumn{1}{c}{Details of the algorithms}                                                                                                                                                 \\ \hline
			Fixed-share experts \cite{105}                                                                                                                                        & Delay-sensitive networks                                                                      & \begin{tabular}[c]{@{}l@{}}Employ the experts framework to predict the RTT and then \\ adjust the network environment to improve the goodput\end{tabular}                                      \\ \hline
			\begin{tabular}[c]{@{}c@{}}Fixed-share with exponentially \\ weighted moving average \\ without increasing computational \\ complexity \cite{110}\end{tabular} & Networks with fluctuating time scales                                                          & \begin{tabular}[c]{@{}l@{}}Propose a technique to estimate the RTT in scenarios with \\ diversified RTT.\end{tabular}                                                            \\ \hline
			Bayesian theorem \cite{53}                                                                                                                                            & \begin{tabular}[c]{@{}c@{}}Real-time video applications \\ and wireless networks\end{tabular} & Adapt the sending rate based on the estimated delay                                                                                                                                            \\ \hline
			Linear Regression \cite{36}                                                                                                                                           & Interactive video applications                                                                & \begin{tabular}[c]{@{}l@{}}Build a statistical function between the sending rate and RTT \\ and adjust the sending rate based on the linear regression given \\ the estimated RTT\end{tabular} \\ \hline
		\end{tabular}
	\end{table*}

\subsection{Queue Length Management in Network-assisted Networks}
	
Queue length management is a key focus for network-assisted CC algorithms. There has been a body of research related with the AQM family of ECN techniques. However, the original AQM algorithms detect the current queue length and react to the environment. Some research has shown that the future queue length can be predicted. The prediction process is shown in Figure 3. Moreover, Table IV summarizes some related research. \cite{78} and \cite{79} showed the long-range dependence between previous traffic patterns and future queueing behavior. Multiple supervised learning techniques have been applied including linear minimum mean square error estimation \cite{88}, normalized least mean square algorithm \cite{89}, neural networks \cite{91}\cite{29}, deep belief networks \cite{13} and neural-fuzzy \cite{1.2}.
	
These algorithms share similar features in that they employ the time series of previous traffic as input without considering diverse parameters in the network. As a result, these algorithms leave space for further exploration of dependencies between related parameters and the queue length. 
	
	\begin{table*}[h]
		\centering
		\caption{Supervised Learning: Queue Management in Network-supported CC algorithms}
		\renewcommand{\arraystretch}{1.5}
		\begin{tabular}{c|c|l}
			\hline
			Algorithms                                                                                                        & Scenarios               & \multicolumn{1}{c}{Details of the algorithms}                                                                                                              \\ \hline
			Neural networks \cite{29}, \cite{91}                                            & ATM networks            & Predict the future value of the traffic based on the past traffic flows                                                                           \\ \hline
			Neural-fuzzy \cite{1.2}                                                                          & ATM networks            & \begin{tabular}[c]{@{}l@{}}Use the estimated average queue length to calculate loss and then control the \\ sending rate\end{tabular}                       \\ \hline
			\begin{tabular}[c]{@{}c@{}}Linear minimum mean square error \\ estimation \cite{88}\end{tabular} & Networks supporting AQM & \begin{tabular}[c]{@{}l@{}}Establish a relationship between long-range traffic flows to estimate the future \\ traffic based on past traffic flows\end{tabular} \\ \hline
			Normalized least mean square  \cite{89}                                                          & Networks supporting AQM & Employ adaptive techniques to estimate the instantaneous queue length                                                                                       \\ \hline
			Deep belief networks \cite{13}                                                                    & NDN   & \begin{tabular}[c]{@{}l@{}}Calculate the average queue length based on the prediction of pending interest \\ table entries\end{tabular}                         \\ \hline
		\end{tabular}
	\end{table*}
	
\section{Unsupervised Learning-based Congestion Control Algorithms}
	
In this section, another category of learning-based CC algorithms is presented: unsupervised learning-based CC algorithms. Unsupervised learning techniques are used when the category of data is unknown, and the sample set needs to be clustered according to the similarity between samples in an attempt to minimize the intra-class gap and maximize the inter-class gap. Classic unsupervised learning algorithms include K-means and Expectation Maximization. Compared to supervised learning-based CC algorithms, unsupervised learning-based CC algorithms are not widely used. They are mainly used to cluster loss and delay characteristics. 
	
\subsection{Congestion Detection in End-to-end Congestion Control Algorithms}  
	
\subsubsection{Loss Clustering}
	
In networking, unsupervised learning techniques are used to cluster loss into several groups and allocate resources for each group to achieve CC as shown in Figure 4. A detailed summary is shown in Table V.
	
In \cite{75}, the packet delay variations reflect the available bandwidth and loss types. Therefore, loss-delay pairs are used to cluster the loss in networks. In \cite{81} and \cite{82}, loss-delay information is utilized. When a packet is lost, it will be marked and tagged with the RTT value. Based on the RTT distribution, these losses can be clustered into two groups: wireless losses and congestion losses. The simulation shows that congestion losses have a higher mean value of RTT while wireless losses have a lower mean and higher variation for RTT. In \cite{87}, the expectation maximization clustering technique is used to cluster losses into contention losses and congestion losses in OBS.
	
Unsupervised learning techniques are useful for training but on their own, they cannot meet the demands of complex networks. Compared with supervised learning techniques, unsupervised learning methods are relatively basic, and are mostly used to represent state spaces \cite{142} and deal with data aggregation \cite{143}. Therefore, research based on this approach is limited.  
	
\subsubsection{Delay Prediction}
	
There are only a limited number of unsupervised learning-based CC algorithms suitable for delay prediction because of the high processing demands for delay calculation. Typical algorithms such as k-means  \cite{21} and the associated mechanisms are presented in Figure 5 and Table VI. Data such as the message size, validity of messages, distance between vehicles and RUSs and the type of message is divided into different groups and the lowest delay in each group is selected as the communication parameter for each cluster. Based on the communication parameter, a specific sending rate will be assigned to each cluster. Therefore, based on the measurement of delay, CC can be achieved.
	
Based on delay features of the network states, clustering is achievable, however, given dynamic and diverse network environments, unsupervised learning techniques are not so well suited compared to supervised learning algorithms. 
	
	\begin{table*}[h]
		\centering
		\caption{Unsupervised Learning: Loss Clustering in End-to-end CC algorithms}
		\renewcommand{\arraystretch}{1.5}
		\begin{tabular}{c|c|l}
			\hline
			Algorithms                                                                  & Scenarios                        & \multicolumn{1}{c}{Details of the algorithms}                                                                                                              \\ \hline
			Hidden Markov Models \cite{81}, \cite{82} & Wired/wireless networks          & \begin{tabular}[c]{@{}l@{}}Uses delay-loss pairs to cluster  data into several groups and assign \\ the specific sending rate for each group\end{tabular} \\ \hline
			Expectation Maximization Clustering \cite{87}              & Optical burst switching networks & \begin{tabular}[c]{@{}l@{}}Cluster  loss into contention loss and congestion loss and adjust the \\ environment separately\end{tabular}                  \\ \hline
		\end{tabular}
	\end{table*}
	
	\begin{table*}[h]
		\centering
		\caption{Unsupervised Learning: Delay Clustering in End-to-end CC algorithms}
		\renewcommand{\arraystretch}{1.5}
		\begin{tabular}{c|c|l}
			\hline
			Algorithms                         & Scenarios                 & \multicolumn{1}{c}{Details of the algorithms}                                                                                                                                                                                                                        \\ \hline
			K-means \cite{21} & Vehicular ad hoc networks & \begin{tabular}[c]{@{}l@{}}Cluster the data into groups based on message size, validity of messages, distance  \\ between vehicles and RSUs, types of message and direction of message sender and assign a sending \\ rate for  each cluster\end{tabular} \\ \hline
		\end{tabular}
	\end{table*}

\begin{figure}[h]
	\centering  
	\includegraphics[height=4.7cm,width=9cm]{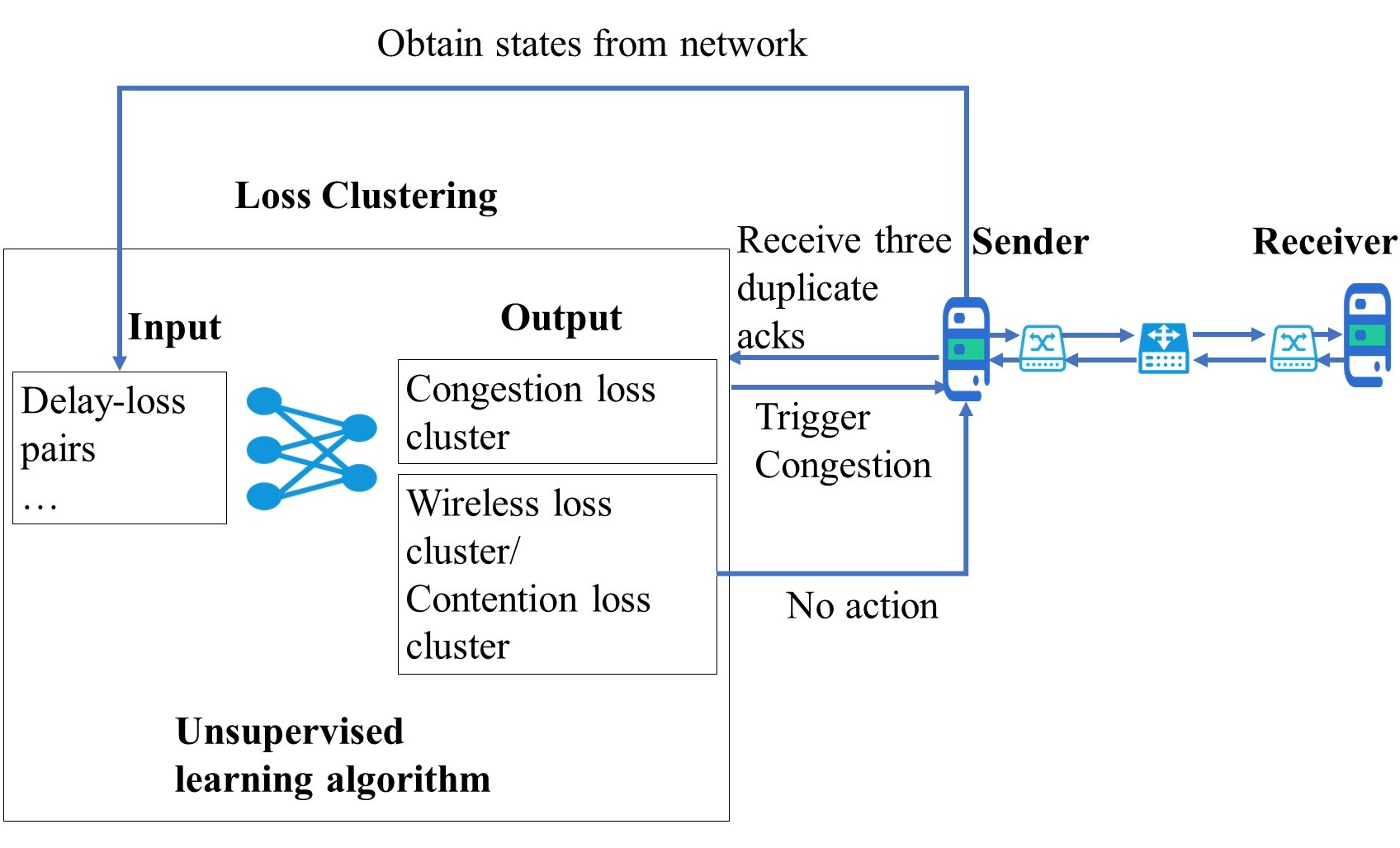}
	\caption{\textbf{Loss Clustering based on Unsupervised Learning Algorithms}}  
	\label{4}  
\end{figure} 

\begin{figure}[h]
	\centering  
	\includegraphics[height=4.7cm,width=9cm]{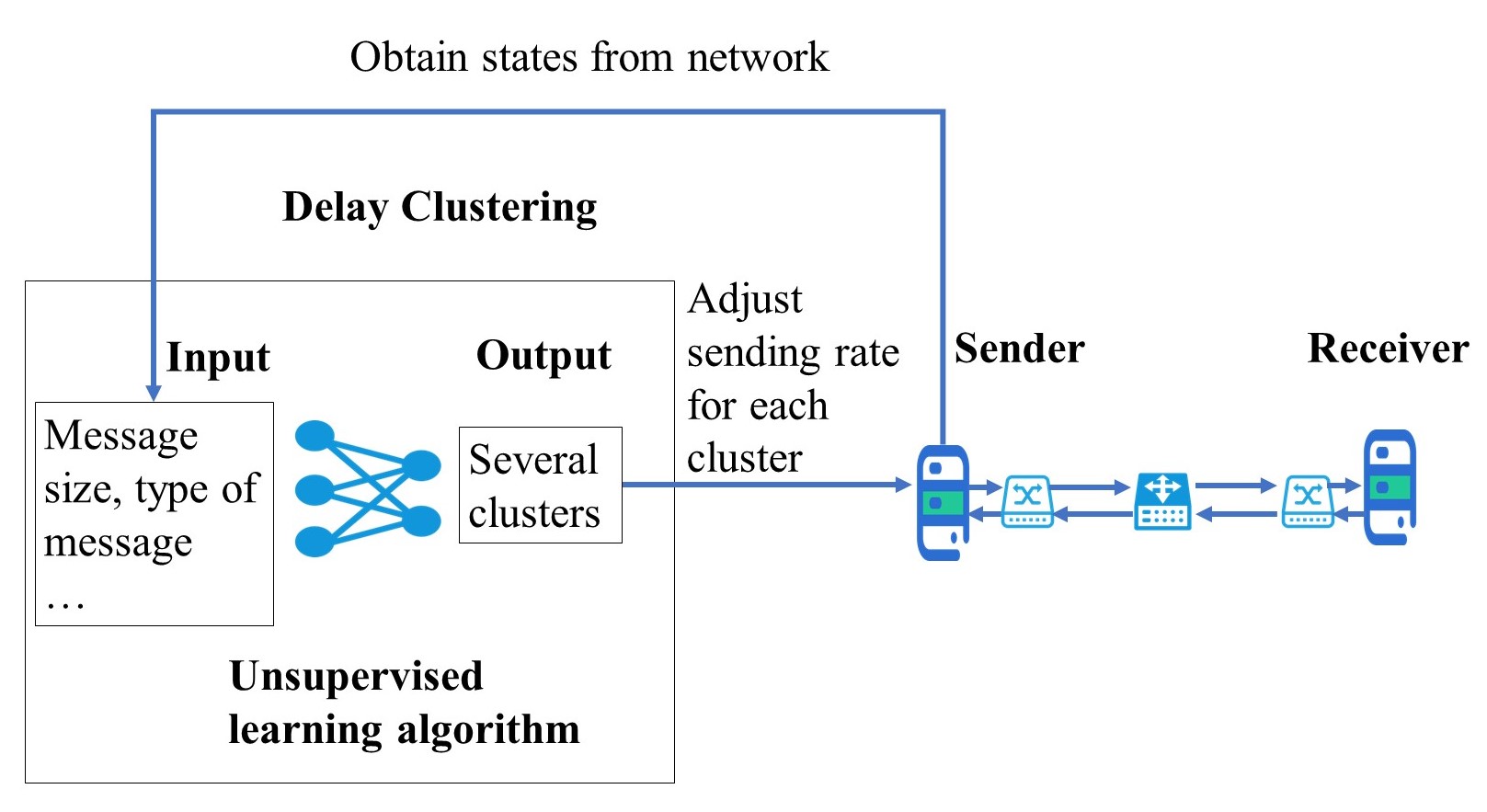}
	\caption{\textbf{Delay Clustering based on Unsupervised Learning Algorithms}}  
	\label{5}  
\end{figure} 
	
\section{RL-based Congestion Control algorithms}

RL algorithms typically include a value function and a policy function. The value function is responsible for measuring the value of specific actions given the network state, to determine if a given  action can be chosen. The policy function is used to choose the action based on a given set of rules. In a given iteration, the system chooses an action based on the policy and the system provides feedback. The value function then calculates the value of the action and updates it accordingly. Based on different mechanisms, RL algorithms are divided into value-based schemes and policy-based schemes. Typical value-based schemes include Q Learning and DQL. Typical policy-based schemes include Policy-Gradient, Actor-Critic (AC), Advantage Actor-Critic (A2C) and Asynchronous Advantage Actor-Critic (A3C). The difference between value-based schemes and policy-based schemes is that policy-based schemes estimate the policy for actions and whether they can satisfy  scenarios with different actions, while value-based schemes predict the value of actions directly. As a consequence they are only suitable for the small set of actions. RL algorithms can be applied in specific networks to improve the efficiency of CC. 

Amongst the different learning-based CC algorithms, RL has gained the most attention. Different to supervised learning methods, RL algorithms monitor the status of environment continuously and react to the environment to optimize a utility function, which leverages the information from the environment. Therefore, RL algorithms are more suitable to  variable and unstable network environments. Two main trends are related with this kind of network. First, ubiquitous applications in data centers and cloud computing require efficient CC algorithms to deal with complicated network topologies \cite{6}. In this context, reliability can be extremely important given the variances that can appear in the system. RL algorithms  adapt to the errors in a timely manner based on learning from the environment. Second, mobile devices such as smart phones, often connect to wireless networks including WIFI and 4G cellular in an ad hoc fashion. As such, more flexible network topologies and diversified flows are a major challenge \cite{3}. Traditional ML approaches are not dynamic enough to cope with diverse network environments based on trained models, unlike RL algorithms. These two trends are driving RL-based CC algorithms. In  RL-based CC algorithms, RL are used to update CWND based on different scenarios in end-to-end networks and to manage the queue length in network-assisted environments.
	
\subsection{Window Updating in End-to-End Networks}  
	
Compared to supervised learning and unsupervised learning techniques, RL algorithms are more responsive to environment changes. Instead of predicting congestion loss and delay as with supervised and unsupervised learning-based CC algorithms, RL-based CC algorithms learn the CC rules directly based on different environment information. Since RL algorithms can incorporate real-time network conditions and define actions accordingly, real-time control is possible in RL algorithms.  
	
Various explorations have focused on RL-based CC algorithms that use RL to update CWND for specific scenarios. The mechanism of RL-based CC algorithms are shown in Figure 6 and the summary is shown in Table VII.
	
ATM is a typical network suitable for RL-based CC algorithms. ATM networks are classic networks that support multi-media applications. For different multimedia traffic, ATM offers different QoS such as cell loss rate and delay. However in ATM, highly time-varying traffic patterns can increase the uncertainty of network traffic. Moreover, the small cell transmission time and low buffer sizes in ATM networks require more adaptive and high responsive CC algorithms. In \cite{1.1}, an AC algorithm is applied to deal with these problems. In the proposed CC algorithm, AC focuses on the performance function based on the cell loss rate and voice quality. In each step, the algorithm measures the action according to the performance. In this way, different traffic patterns are connected with corresponding actions. Simulation results show that the cell loss rate is low and voice quality is maintained. 
	
Software Defined Networks (SDNs) provide  a new architecture for future networks that separate the forwarding  and control planes. The control plane has the ability to manage the overall network centrally. Efficient CC algorithms are essential for SDNs. In \cite{6}, Q learning is used to tackle such advanced networks. The trained algorithm show that higher link utilization can be achieved.
	
Named Data Networking (NDN) is an emerging future network architecture as well. The main characteristic of NDN is connectionless, providing content perceptibility and in-network caching. Typical applications of NDN are mobile  and real-time communications. Therefore, CC algorithms are expected to cope with diverse and dynamic content. In \cite{52}, the deep RL algorithm considers the diversity of different content and adds a prefix when requesting content into the network. Therefore, the variety of content is considered when a given action is taken.
	
Satellite communication networks are dynamic and have time-varying flows. High bandwidth and high elasticity are key features. Video streaming is one representative application. In satellite communication networks, frequent satellite handover can be a severe problem, which may result in routing failures, packet blocking and channel quality impacts. To deal with these problems, \cite{131} employs DDPG to design a multi-path TCP. By measuring the re-transmission rate of each sub-flow, the RTT and ACK number are considered and the algorithm degrades the possibility of handover. 
	
Internet of Things (IoT) is a product of rapidly evolving wireless technology. Some core features of IoT are local computation, high variability of use and potential computational demands. In \cite{99}, Q learning was used to satisfy diverse IoT networks with reduced computational needs with strong learning capabilities. The proposed algorithm  showed that the adjustment action was suitable for real-time processors and memory demands of IoT environments.
	
Wired networks are not typical scenarios in learning-based CC field. Wired networks are relatively stable compared to wireless networks. Of course, some research covers this scenario as well e.g. \cite{20}. In \cite{20}, high bandwidth and under-buffered bottleneck links were taken into consideration, as typical features of wired networks. The states of the algorithm included multiple parameters such as packet inter-sending time and inter-arrival time of input ACK reflecting the information of the current available buffer information. Therefore, the algorithm achieved a better balance between throughput and delay.
	
Wireless networks are a research hot-spot for learning-based CC algorithms especially Ad hoc Wireless Networks (AWNs). AWNs are a collection of mobile wireless nodes without any fixed infrastructure. Therefore, AWNs have constrained resources, limited processing and unpredictable mobility. They are also highly dynamic. In \cite{96}, Finite Action-set Learning Automata, a learning automata whose unique feature is learning the network state faster with reduced information and negligible computational requirements, contains a finite number of actions. The algorithm takes effect in learning the dynamic wireless environment with limited consumed resources. While in \cite{97}, Continuous Action-set Learning Automata was applied in AWNs. The discretization of Finite Action-set Learning Automata may not be proper in all situations, e.g. the discretization can be too coarse or too fine-grained. Therefore, Continuous Action-set Learning Automata was introduced to deal with an infinite number of actions. It maintains an action probability distribution. The advanced algorithm achieves better performance. Of course, more computational and training resources are consumed. Moreover, in \cite{98}, Q learning combined with a grey model was used to predict throughput and performance of CC algorithms in AWNs. Due to the real-time evaluation of throughput, the algorithm adapts to the dynamic environments better.
	
The RL-based CC algorithms above focus on single scenarios, however there are some RL-based designed for more complex (multiple) network scenarios. For instance, \cite{2}, \cite{8}, \cite{17} and \cite{12} propose an AC algorithm to deal with congestion problems in networks with time-varying flows. In \cite{3}, the RL-based CC algorithms are used in networks with sparse rewards such as video games, while in \cite{4}, the scenario focuses on continuous, large state-action spaces.  
	
From the above, it can be seen that RL-based CC algorithms can satisfy diverse network scenarios with high adaptability and strong flexibility. However, there are some limitations. For instance, convergence is very hard to guarantee for continuous tasks and complex algorithms. In addition, state abstraction is challenging. Current algorithms require significant storage to store states and actions and demand considerable memory resources. Moreover, their computational complexity is relatively high. As a result, though RL algorithms show strong learning capabilities, realistic applications require further exploration due to the engineering issues identified.     

\begin{figure}[h]
	\centering  
	\includegraphics[height=4.7cm,width=9cm]{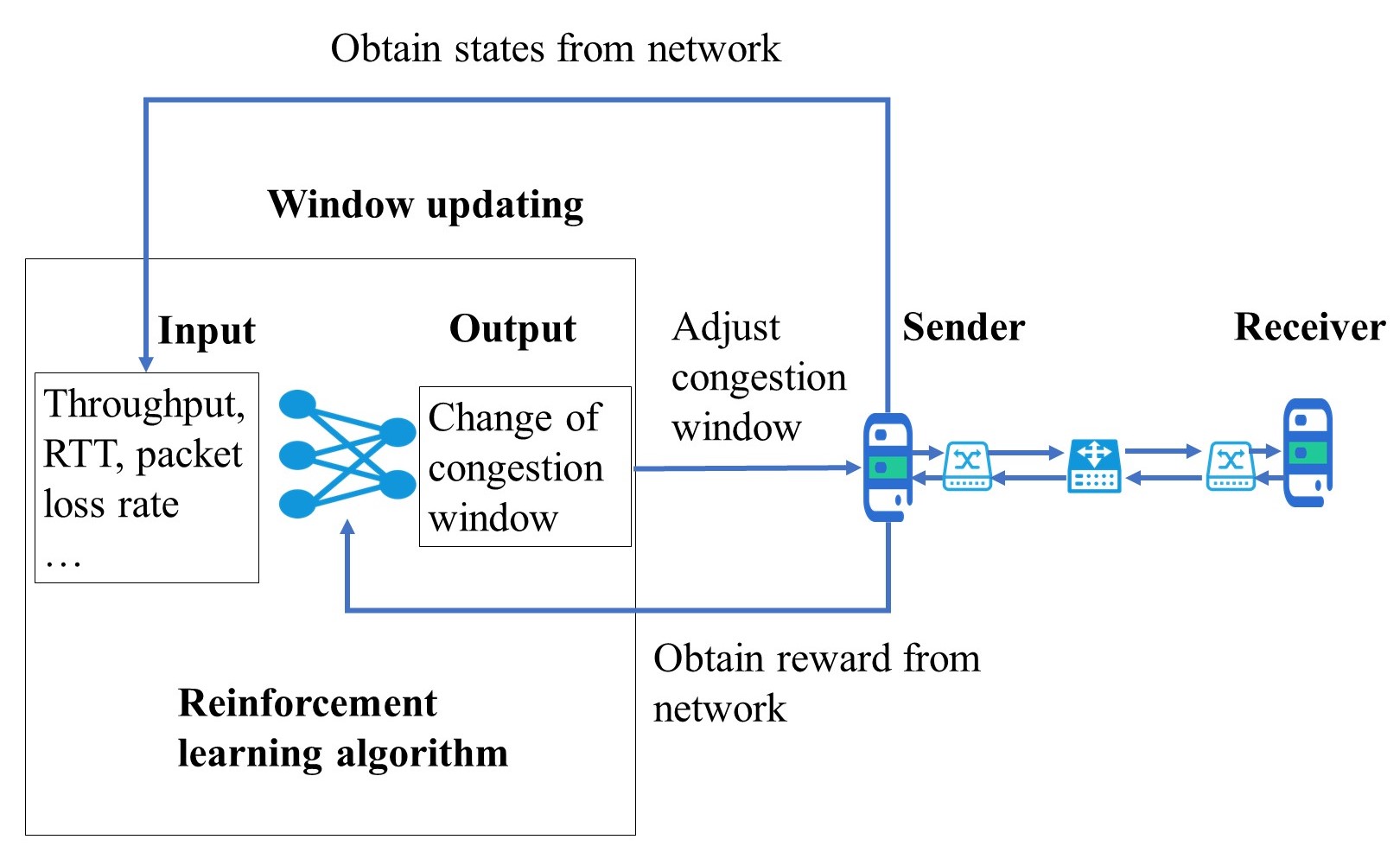}
	\caption{\textbf{Window Updating based on RL Algorithms}}  
	\label{6}  
\end{figure}   

\begin{figure}[h]
	\centering  
	\includegraphics[height=4.7cm,width=9cm]{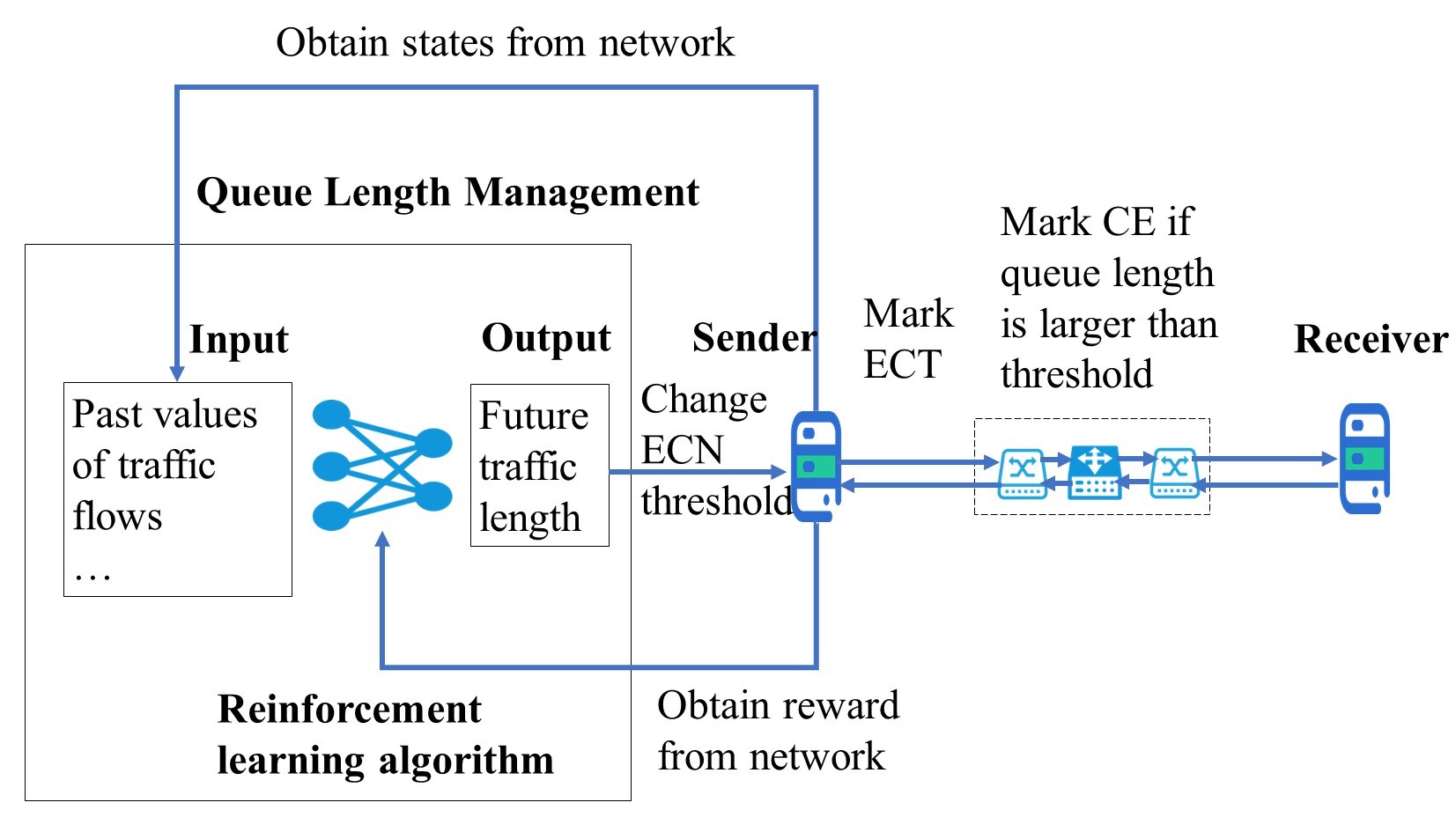}
	\caption{\textbf{Queue Length Management based on RL Algorithms}}  
	\label{7}  
\end{figure} 
	
	\begin{table*}[]
		\centering
		\caption{RL: Window Updating in End-to-End CC Algorithms}
		\renewcommand{\arraystretch}{1.5}
		\begin{tabular}{c|c|l}
			\hline
			Algorithms                                                          & Scenarios                                                                                                                                        & \multicolumn{1}{c}{Details of the Algorithms}                                                                                                                                                                                                      \\ \hline
			AC \cite{1.1}                                      & ATM networks                                                                                                                                     & \begin{tabular}[c]{@{}l@{}}Employ the actor critic algorithm to minimize packet loss \\ rate and preserve  video/voice quality\end{tabular}                                                                                                  \\ \hline
			Q learning and Sarsa \cite{6}                      & SDN                                                                                                                                              & \begin{tabular}[c]{@{}l@{}}Train an off-policy method based Q learning and an \\ online-policy method based on Sarsa to control congestion. \\ Both algorithms achieve good link utilization\end{tabular}                                  \\ \hline
			DQL \cite{52}                                      & NDN                                                                                                                            & \begin{tabular}[c]{@{}l@{}}Learn an optimal CC policy by taking the diversified contents \\ in NDN\end{tabular}                                                                                                   \\ \hline
			DDPG \cite{131}                                     & MPTCP in satellites communications                                                                                                               & \begin{tabular}[c]{@{}l@{}}Present an intelligent algorithm to improve the performance \\ of low earth orbit satellite communications\end{tabular}                                                                                                 \\ \hline
			Fuzzy Kanerva-based Q Learning \cite{99}           & IoT                                                                                                                                              & \begin{tabular}[c]{@{}l@{}}Reduce the amount of memory needed to store the algorithm \\ history to support  larger state spaces and  action spaces\end{tabular}                                                                                     \\ \hline
			Q learning \cite{20}                               & \begin{tabular}[c]{@{}c@{}}Wired networks with under-buffered \\ bottleneck links\end{tabular}                                                   & \begin{tabular}[c]{@{}l@{}}Input acknowledgement inter-arrival time, packet inter-sending \\ time, the ratio of the current RTT, minimum RTT, the slow start \\ threshold and CWND size to get adjustment information\end{tabular} \\ \hline
			Finite Action-set Learning Automata \cite{96}      & AWNs                                                                                                                        & \begin{tabular}[c]{@{}l@{}}Input the data including the inter-arrival times of ACKs and \\ duplicate packets and output the window size\end{tabular}                                                                                                \\ \hline
			Continuous action-set learning automata \cite{97}  & AWNs                                                                                                                     & Maintain an action probability distribution                                                                                                                                                                                                         \\ \hline
			Q learning \cite{98}                               & AWNs                                                                                                                & \begin{tabular}[c]{@{}l@{}}Take throughput and RTT into consideration when projecting \\ the state spaces to  action spaces\end{tabular}                                                                                                         \\ \hline
			DQL \cite{9}                                       & Wireless networks                                                                                                                                & \begin{tabular}[c]{@{}l@{}}Input the states consisting of CWND, RTT and the \\ inter-arrival time and then output the sending rate\end{tabular}                                                                                        \\ \hline
			Continuous action-set learning automata \cite{100} & \begin{tabular}[c]{@{}c@{}}Wireless networks: Multi-hop, \\ single-hop such as  wireless LANs, \\ cellular, and satellites networks\end{tabular} & Maintain an action probability distribution                                                                                                                                                                                                         \\ \hline
			AC \cite{2}                                        & Network with time-varying flows                                                                                                                  & \begin{tabular}[c]{@{}l@{}}Design a multi-agent congestion controller based on the \\ actor-critic framework\end{tabular}  
			\\ \hline
			AC \cite{8}                                        & Network with time-varying flows                                                                                                                  & \begin{tabular}[c]{@{}l@{}}AC algorithm is applied in LSTM-based representation networks, \\which shows effectiveness and superiority compared with \\well-known MPTCP CC algorithms such as wVegas\end{tabular}                                                                                                                              \\ \hline
			A3C \cite{3}                                       & \begin{tabular}[c]{@{}c@{}}Task with sparse reward such as video \\ games\end{tabular}                                                           & \begin{tabular}[c]{@{}l@{}}Propose a partial action learning method which supports delayed \\ and partial rewards\end{tabular}                                                                                                                      \\ \hline
			Q learning \cite{4}                                & Continuous or large state-action space                                                                                                           & \begin{tabular}[c]{@{}l@{}}Abstract the state space and action space based on Kanerva \\ coding\end{tabular}                                                                                                                                    \\ \hline
			PPO \cite{5}                                       & \begin{tabular}[c]{@{}c@{}}Internet services such as live video, \\ virtual reality and internet-of-things\end{tabular}                          & \begin{tabular}[c]{@{}l@{}}Detect network and data patterns such as latency to get the \\ necessary adjustment\end{tabular}                                                                                                                                       \\ \hline
			Q learning \cite{17}                               & Dynamic networking                                                                                                                               & \begin{tabular}[c]{@{}l@{}}Detect the average packet arrival interval, average ACK \\ interval and average RTT to adjust the CWND size\end{tabular}                                                                                        \\ \hline
			A3C \cite{12}                                      & Network with diversified flow size                                                                                                               & \begin{tabular}[c]{@{}l@{}}Employ the RL algorithm to configure the initial window \\ and CC policy\end{tabular}                                                                                                    \\ \hline
		\end{tabular}
	\end{table*}
	
\subsection{Queue Length Management in Network-assisted Networks}

For the queue length management of RL-based CC algorithms, RL is used to manage the queue length based on the current state as shown in Figure 7 and Table VIII. In queue management, Proportional Integral Derivative (PID) is the most commonly applied RL technique. In \cite{93,94,95}, PID is used to maintain the queue length given the target threshold by calculating the dropping probability. Congestion notification is used to control the queue length as well in \cite{111}. The proposed algorithm employs Q learning to properly utilize buffer size in disruption tolerant networks. With the objective of maximizing the link utilization based on the queue length management, \cite{90} and \cite{92} use loading information to optimize router decisions relying on RL algorithms.
	
Compared with window updating for end-to-end networks, the queue length management for network-assisted CC algorithms requires more computational resources because multiple nodes can be used to control congestion such as routers in network-assisted networks. Therefore, it may be a burden for the network to support RL-based CC algorithms given the larger state space and high computational complexity. In addition, current queue length management based on RL techniques only cover limited state parameters such as the past queue length and buffer size. However, more parameters are required  to improve the performance of RL-based CC algorithms. 
	
	\begin{table*}[]
		\centering
		\caption{RL: Queue Management in Network-supported CC algorithms}
		\renewcommand{\arraystretch}{1.5}
		\begin{tabular}{c|c|l}
			\hline
			Algorithms                                               & Scenarios                    & \multicolumn{1}{c}{Details of the Algorithms}                                                                                                                                  \\ \hline
			PID controller \cite{93}                & Networks supporting AQM      & Employ PID to adapt the parameters in networks by stabilizing the router queue length                                                                                             \\ \hline
			Adaptive neuron PID \cite{94}           & Networks supporting AQM      & \begin{tabular}[c]{@{}l@{}}Given different traffic loading, scenarios, RTTs, bottleneck link capacities, maintain \\ the queue length around a target queue length\end{tabular} \\ \hline
			Q learning \cite{92}                    & Networks supporting AQM      & Use RL to optimize router decisions based on traffic history                                                                                                \\ \hline
			Neuron RL \cite{90}                     & Networks supporting AQM      & Control the queue length and maximize the link utilization based on queue management                                                                                            \\ \hline
			Neural network PID controller \cite{95} & Networks supporting AQM      & Based on the learning rate, calculate the dropping probability                                                                                                                  \\ \hline
			Q learning \cite{111}                   & Disruption tolerant networks & Employ the congestion state to support congestion notifications                                                                                                                       \\ \hline
		\end{tabular}
	\end{table*}
	
\section{Simulation Setup}
	
In this section, we introduce the simulation setup for RL-based CC algorithms as representatives of learning-based CC approaches. We conduct  experiments based on realistic network environments with challenges caused by large delay and their high complexity. We perform experiments based on the NS3 platform and explore the performances of RL-based CC algorithms and  traditional CC algorithms. In the NS3 platform, the computational process related to the RL algorithms is separated from data transmission in pipeline. As a result, the computational complexity of RL algorithms has no impact on network communications.

In the following sections, we compare algorithms, performance metrics and network environments.
	
\subsection{Compared Algorithms}

In the simulation, three RL algorithms are chosen: DQL, DDPG and PPO, as typical examples of RL algorithms. Generally, DQL is the simplest among these three algorithms, hence it is suitable for relatively simple environments. DDPG and PPO have stronger learning capabilities, and hence they can be applied in more complex scenarios. Considering the limited complexity of our network environment, these three algorithms are expected to perform similarly. To compare them with a benchmark algorithm, NewReno was selected, which is a classic traditional CC algorithm and is the default CC algorithm of NS3 as well. These four algorithms are summarized in Table IX.
	
	\begin{table*}[h]
		\centering
		\caption{Compared CC algorithms}
		\renewcommand{\arraystretch}{1.5}
		\begin{tabular}{c|l|l|l|l}
			\hline
			Techniques & \multicolumn{1}{c|}{Applied scenarios}                                                                                        & \multicolumn{1}{c|}{Mechanism}                                                                                                                                                                                                                                   & \multicolumn{1}{c|}{Advantage}                                                                                      & \multicolumn{1}{c}{Limitation}                                                                                                                                                                                                                   \\ \hline
			DQL        & \begin{tabular}[c]{@{}l@{}}Wired/wireless network, \\ NDN \end{tabular}                                      & \begin{tabular}[c]{@{}l@{}}Input state and output action values \\ based on neural networks\end{tabular}                                                                                                                                                      & \begin{tabular}[c]{@{}l@{}}Have the ability to solve large-scale \\ RL problems\end{tabular}                        & \begin{tabular}[c]{@{}l@{}}Cannot guarantee convergence \\ of networks\end{tabular}                                                                                                                                                              \\ \hline
			DDPG       & \begin{tabular}[c]{@{}l@{}}MPTCP in satellites\\ communications\end{tabular}                                                  & \begin{tabular}[c]{@{}l@{}}Combine DQL and AC algorithms, \\ consisting of two Actor networks and \\ two Critic networks. In addition, they \\ adopt a deterministic policy in each step\end{tabular}                                                              & \begin{tabular}[c]{@{}l@{}}Obtain good performance and converge \\ quickly in continuous action spaces\end{tabular} & \begin{tabular}[c]{@{}l@{}}Not suitable for random \\ environments\end{tabular}                                                                                                                                                                   \\ \hline
			PPO        & \begin{tabular}[c]{@{}l@{}}Internet services such as \\ live video, virtual\\  reality and \\ internet-of-things\end{tabular} & \begin{tabular}[c]{@{}l@{}}Propose a new objective function that \\ can be updated in small batches with \\ multiple training steps, solving \\ the problem that the step size in the\\ policy gradient algorithm is difficult \\ to determine\end{tabular} & \begin{tabular}[c]{@{}l@{}}Guarantee the convergence and \\ performance\end{tabular}                                & \begin{tabular}[c]{@{}l@{}}The speed of policy updating is \\ related with the direction of policy \\ gradient which ignores the space \\ structure of policy parameters. \\ Therefore, the speed of training \\ a policy may be slow\end{tabular} \\ \hline
			NewReno    & Wired networks                                                                                                                & \begin{tabular}[c]{@{}l@{}}Consist of four parts: slow start, \\ congestion avoidance, re-transmission \\ and fast recovery\end{tabular}                                                                                                                           & \begin{tabular}[c]{@{}l@{}}Avoid  inefficiency of slow start \\ processes and guaranteed throughput\end{tabular}  & \begin{tabular}[c]{@{}l@{}}Cannot proactively determine \\ congestion and predict packet loss\end{tabular}                                                                                                                                       \\ \hline
		\end{tabular}
	\end{table*}

\subsubsection{DQL-based Congestion Control Algorithms}
	
Different from Q Learning or Sarsa which considers the state as a discrete finite set, DQL can deal with large scale problems. In the DQL algorithm, the value function is expressed by neural networks such as Convolutional Neural Networks (CNN), Recurrent Neural Networks (RNN) and Long Short-Term Memory (LSTM). For the value function of DQL algorithms, there are two main methods. One method uses the state and  action as inputs, to get the action value as the output from the neural networks. Another method is where the state is the input, and actions and related action values are the outputs. These two methods imply that the action space provides a finite number of discrete actions. Because DQL approximates the value function through the neural network, DQL can solve large-scale problems. However, DQL has a problem, in that it does not necessarily guarantee the convergence of the Q network. As such, it may not be able to get the Q network parameters after convergence. This will result in a poorly trained model. However, in the network field, DQL still exhibits high performance especially when dealing with complex networks.
	
TCP-Drinc is an efficient RL-based CC algorithm which uses a deep CNN concatenated with a LSTM network to learn from historical data. It determines the next action and then adjusts the CWND size. LSTM is suitable for processing and predicting important events with very long intervals and delays in time series. In Drinc, LSTM is utilized to handle auto-correlations within the time-series introduced by delays and related information. Therefore, the DQL framework is robust and has a better learning capability. Moreover, Drinc is designed for multi-agent networks and can deal with varying network conditions \cite{9}. 

DQL is relatively simple compared with other Deep RL and has the capability to deal with relatively simple networks. Except for the convergence issue, DQL is promising because the model is lighter.
	
\subsubsection{DDPG-based Congestion Control Algorithms}
	
DDPG is an optimized version of the AC algorithm, which converges quickly and performs well. To better understand DDPG algorithms, AC algorithms are introduced.
	
The AC algorithm is based on the policy gradient method, which is a policy-based RL algorithm. For value-based RL algorithms such as Q Learning and DQL, these methods generally only deal with discrete actions and hence they cannot handle continuous actions nor solve stochastic problems. Therefore new approaches to cover these scenarios such as policy-based methods are required. In value-based methods, the value function is approximated and used to calculate the action value based on the input of the state and the associated action. In policy-based methods, the algorithm adopts a similar approach but approximates the strategy instead.

AC algorithms combine policy-based methods and value-based methods. The actor part is used to approximate the policy function and is responsible for generating actions that interact with the environment. Given a policy function $\prod_{\theta}(s,a)$, the critic part is used to approximate the value function and evaluate the performance of the actor of the next stage. The most commonly used policy function is the \textit{Softmax} strategy function. It is mainly used in discrete spaces. The \textit{Softmax} strategy uses a linear combination of characteristics ($\varphi$(s,a)) describing the state and the parameter $\theta$ to weigh the probability of a behavior occurring. The function is given as:
	\begin{equation}
	\prod_{\theta}(s,a)={\displaystyle\frac{e^{{\varphi}(s,a)^T{\theta}}}{{\sum}_{b}e^{{\varphi}(s,b)^T{\theta}}}}
	\end{equation}

The corresponding score function is obtained by derivation and is given as:
	\begin{equation}
	{\bigtriangledown_\theta}log{\prod_{\theta}(s,a)}={\varphi}(s,a)-E_{\prod_{\theta}}{\varphi}(s,a)
	\end{equation}
	
	The function of parameter for updating for the policy is $\theta$:
	\begin{equation}
	{\theta}={{\theta}+{\alpha}{{\bigtriangledown_\theta}log{prod_{\theta}(s_t,a_t)}v_t}}
	\end{equation}
	
where $v_t$ is the Q value given state $s_t$ and action $a_t$. The critic unit, refers to the DQL-based CC algorithm which employs Q learning as a critic and obtains the action value, before updating the parameter of Q learning. 
	
AC takes advantage of both mainstream RL algorithms, but they can be  difficult to converge since there are two neural networks which are related to each other and both require updating of the gradient. 
	
The early versions of AC-based CC algorithms were designed for routing-based algorithms. In \cite{1.1}, the proposed algorithm was designed for CC for multi-media traffic in ATM networks through deep neural networks. The result showed that the presented algorithm achieved a high voice/video quality by reducing losses and delays \cite{1}. Later an AC-based algorithm was used as an effective technique for multi-path CC. Similar to DQL-based algorithms, the proposed AC-based algorithm integrated LSTM to represent the state-action space. Simulations showed that the presented algorithm was flexible for networks with continuous action spaces and performed favorably to traditional CC algorithms \cite{8}.
	
AC-based CC algorithms offer advanced explorations which are not always robust. The performance of this kind of algorithms depends on the interaction of the two neural networks. This requires further research to guarantee their convergence and overall efficiency.
	
DDPG is another category of RL algorithm to deal with the convergence issue of AC. It employs experience reply and double networks. On the one hand, compared with traditional policy gradient algorithms, DDPG outputs a deterministic policy instead of a random policy. Traditional policy gradient algorithms calculate the gradient based on the stochastic strategy gradient. On the other hand, DDPG adopts double actor networks and double critic networks. For the double actor networks, one is responsible for updating policy parameters and the other selects the next action based on sampling from experience replay data sets. For the double critic networks, one updates the parameters related to the Q value and the other calculates the Q value. In satellites communications, a DDPG-based algorithm was designed to deal with multi-path CC problems and achieved a high degree of effectiveness \cite{131}.
	
As shown above, compared with DQL, DDPG has stronger capability to train models in more complex environments. However, DDPG exhibits other problems which make it unsuitable for random environments. In addition, training DDPG models can be more difficult. 
	
\subsubsection{PPO-based Congestion Control Algorithms}
PPO is a deep RL algorithm based on AC schemes. PPO is used to solve  problems where  the traditional policy gradient method is not good enough to determine the learning rate or step size. If the step size is too large, the policy will keep moving and will not converge. However, if the step size is too small, it is time-consuming. To deal with this problem, PPO limits the updating range of new policies by using the ratio between the new and old policy, making the policy gradient less sensitive to slightly larger step sizes. To achieve this, PPO uses an adaptive penalty to control the change in policy. In this way, PPO provides an optimized AC algorithm as well as improving the efficiency of convergence.
	
To adapt to the variable network conditions, such as changeable link flows and end to-end latency, PPO is presented as a RL-guided CC algorithm  \cite{56}. The designed algorithm, Aurora takes advantage of PPO to generate efficient policies for learning and ensuring that the learning procedure is stable. Simulations show that the proposed algorithm outperforms traditional CC algorithms in different contexts by generating optimal policies. 

PPO has proven to be an outstanding deep RL method and the combination with CC shows the potential of PPO in a wide array of network applications. However, there exist some challenges such as the speed of training a policy related to the parameter structures. As a result, the training efficiency of PPO can be a major issue.
	
\subsubsection{NewReno}
	
NewReno is a loss-based CC algorithm based on Reno. It offers a slow start, congestion avoidance, re-transmission and fast recovery. Compared with classic CC algorithms, NewReno modifies the fast recovery part. In the fast recovery of Reno, the sender quits the fast recovery state after receiving a new ACK. In  NewReno, it enters the fast recovery state only after all messages have been acknowledged. Therefore, TCP distinguishes situations such as losing multiple packets in one congestion from  multiple congestion scenarios, and then halves the CWND only once after each congestion occurs, thereby improving the robustness and throughput. In our experiments, NewReno algorithm is used as the representative of traditional CC algorithms. 

\subsection{Performance Metrics}

Based on the literature, the network cares about several critical parameters including throughput, RTT and packet loss rate. Therefore, in our experiments, our performance metrics focus on throughput, RTT and packet loss rate. Throughput counts the amount of data successfully transmitted in a given unit of time,  measured in Mbps. RTT measures the data transfer time from the sender to the receiver based on the average RTT in seconds. Packet loss rate calculates the ratio of packet loss in a given time interval. 

\subsection{Network Environment}
	
\subsubsection{Internet}
All simulations employ the same network topology, comprising the same dumbbell topology with the same access delay and bandwidth. To simulate different network environments, the bottleneck bandwidth and bottleneck delays are varied. 

Based on previous research, learning-based CC algorithms are more suitable for high speed networks such as satellite communications networks \cite{23}, ATM networks\cite{29} and networks with time-varying flows \cite{2}. We speculate that learning-based CC algorithms are suitable in networks with high BDP (Bandwidth-delay Product) since they are more aggressive in making use of  higher BDP. The BDP can be a critical parameter to measure the network as it is used to control congestion in BBR as well \cite{127}. Therefore, we design three scenarios as shown in Table X to compare the performance of the NewReno algorithm and the RL-based CC algorithms.

In the scenarios, there are two senders and two receivers in the dumbbell network. The access bandwidth is 1000Mbps and the access delay is 0.01 milliseconds. In our experiments, high BDP and low BDP are relative and not absolute. In scenario I, the BDP is high and the bottleneck bandwidth is high. However, the bottleneck delay is low. In scenario II, the BDP is high, but the bottleneck bandwidth is low and the bottleneck delay is high. In scenario III, the BDP is low, however the bottleneck bandwidth and bottleneck delay are low. 

In scenario I, the bottleneck delay is set to 2.5 milliseconds. The bottleneck bandwidth  changes from 100M to 140M in 5 seconds. More specifically, the bottleneck bandwidth is 100M initially and incremented by 10M each second up to a maximum of 140M. 

In scenario II, the bottleneck delay is set to 25 milliseconds. The bottleneck bandwidth changes from 10M to 50M in 15 seconds. Every three seconds, the bottleneck bandwidth increases by 10M. Because the bottleneck delay is longer, more simulation time is required in scenario II compared to scenario I. This allows to observe the performance of different CC algorithms.

In scenario III, the bottleneck delay is set to 2.5 milliseconds and the bottleneck bandwidth changes from 10M to 50M in 5 seconds, i.e. every  second the bottleneck bandwidth increases by 10M.

\begin{table*}[]
	\centering
	\caption{Simulation Scenarios}
	\renewcommand{\arraystretch}{1.5}
	\begin{tabular}{l|l|l}
		\hline
		Scenarios    & \multicolumn{1}{c|}{Experiment Setting}                                                                                                                                                                                                                              & BDP                       \\ \hline
		Scenario I   & \begin{tabular}[c|]{@{}l@{}}Access bandwidth: 1000M\\   Access delay: 0.01ms\\   Bottleneck bandwidth: changing from 100M to 140M in 5 seconds (bottleneck bandwidth increases by 10M every second)\\   Bottleneck delay: 2.5ms\end{tabular}             & High                      \\ \hline
		Scenario II  & \begin{tabular}[c]{@{}l@{}}Access bandwidth: 1000M\\   Access delay: 0.01ms\\   Bottleneck bandwidth: changing from 10M to 50M in 15 seconds (bottleneck bandwidth increases by 10M every three seconds)\\   Bottleneck delay: 25ms\end{tabular} & \multicolumn{1}{c}{High} \\ \hline
		Scenario III & \begin{tabular}[c]{@{}l@{}}Access bandwidth: 1000M\\   Access delay: 0.01ms\\   Bottleneck bandwidth: changing from 10M to 50M in 5 seconds (bottleneck bandwidth increases by 10M every  second)\\   Bottleneck delay: 2.5ms\end{tabular}               & Low                       \\ \hline
	\end{tabular}
\end{table*}

\subsubsection{States}
States often vary in different research approaches. In DQL-based CC algorithms, states mainly focus on CWND differences, RTT and the inter-arrival time of ACKs \cite{9}. In a multi-agent CC based on AC, states are based on the buffer length and sending rate \cite{2}. In the A3C framework, states are based on throughput, loss and RTT \cite{12}. In self-learning CC algorithms relying on DDPG, states are based on CWND, RTT, ACK and the cumulative rate number of re-transmissions of the sub-flow \cite{54}. While in PPO, the states are designed in three parts: the latency gradient, the latency ratio and the sending ratio \cite{56}. It is clear that there are no guaranteed rules underpinning RL-based CC algorithms. According to previous literature, states are used to tackle two key areas: congestion signals including RTT, loss, ACK, throughput and the parameter used to control congestion such as the CWND size and the sending rate. In CC algorithms, the environment adjusts the sending rate or the CWND size based on the congestion signals.
	
Considering the focus on performance metrics, the states considered here are throughput, RTT, packet loss rate.
	
\subsubsection{Actions}
	
In the simulations, all adjustments are window-based. By adjusting the CWND size,  there are different rules that are applied. In \cite{20}, there are four actions:\textit{ -1, 0, +1, +3}. When the action is \textit{-1}, the CWND will decrease one packet size. In \cite{17}, three actions are designed: \textit{-1, 0, +10}. The increasing action is more aggressive (up to 10). In \cite{21}, the action space is much larger. Seven actions are predefined: \textit{+1, *1.25, *1.5, 0, -1, *0.75, *0.5}. When the action is \textit{*1.25}, the size of new CWND is 1.25 times  the original CWND. In our experiments, we considered four actions: \textit{-1, 0, +1, +3}, as aligned with \cite{20}.
	
\subsubsection{Rewards}
Similar to states, rewards can have different definitions as well. In a DQL-based CC algorithm, the utility function of reward is defined as shown below \cite{21}.
	\begin{equation}
	\begin{split}
	Utility(t)={\alpha}_i*log({throughput}_i(t))-{\beta}_i*{RTT}_i(t)\\-{\gamma}_i*{loss}_i(t)-{\delta}*{reordering}_i(t)
	\end{split}
	\end{equation}
	
In the PPO-based  CC algorithm \cite{56}, the utility function is defined as shown below: 
	\begin{equation}
	Utility=10*throughput-1000*latency-2000*loss
	\end{equation}
	
In a A3C-based algorithm \cite{12}, the utility function is given as: ${log(throughput/RTT)}$. In a DDPG-based algorithm, the utility function is more complicated \cite{54} and given as: 
	\begin{equation}
	Utility={\sum\nolimits}_i({\alpha}{CWND}_t-{\beta}{rtt}_t-{\epsilon}{rta}_t-k{ack}_t)
	\end{equation}
	
To define the reward, the purpose of the simulation should be defined first. Reward is used for feedback of the action given the current state. Using this, it measures the performance of the action. Thus the reward is a reflection of the performance of actions. From the above, the definition of reward covers throughput, delay and packet loss rate. Considering these factors, the reward includes RTT and throughput. The utility function is shown below where the value of the utility reward is based on \cite{21}. The bandwidth in the equation means the bottleneck bandwidth. \textit{MinRTT} means the minimum RTT of the pipeline. \textit{P} is used for the packet loss rate. 
	\begin{equation}
	\begin{split}
	Utility=log(throughput/(bandwidth))\\-log(RTT-MinRTT)+log(1-p)
	\end{split}
	\end{equation}

\section{Simulations} 
	
In this section, we present the results of the simulations of the four algorithms: traditional CC algorithm NewReno and the RL-based CC algorithms, DQL, DDPG and PPO. The simulations were conducted on the NS3 platform. 
	
Based on previous research, the state space considered includes five parameters: throughput, RTT and packet loss rate. The reward function is given as $log(throughput/bandwidth)-log(RTT-MinRTT)+log(1-p)$. The action is used to adjust the CWND once a new ACK arrives. A dumbbell network topology was adopted.
	
	\begin{table*}[]
		\centering
		\caption{Simulation Results}
		\renewcommand{\arraystretch}{1.5}
		\begin{tabular}{c|c|c|c|c|c}
			\hline
			Scenarios    & BDP  & CWND                 & Throughput           & RTT               & Packet Loss Rate \\ \hline
			Scenario I   & High & Substantial Increase & Substantial Increase & Limited Increase  & Limited Increase \\ \hline
			Scenario II  & High & Substantial Increase & Substantial Increase & Limited Increase  & Limited Increase \\ \hline
			Scenario III & Low  & No big difference    & No big difference    & No big difference & Limited Increase \\ \hline
		\end{tabular}
	\end{table*}
	
	\begin{figure*}[htbp]
		\begin{minipage}[t]{0.5\linewidth}
			\vspace{-0.3cm}
			\centering
			\includegraphics[width=9cm,height=4cm]{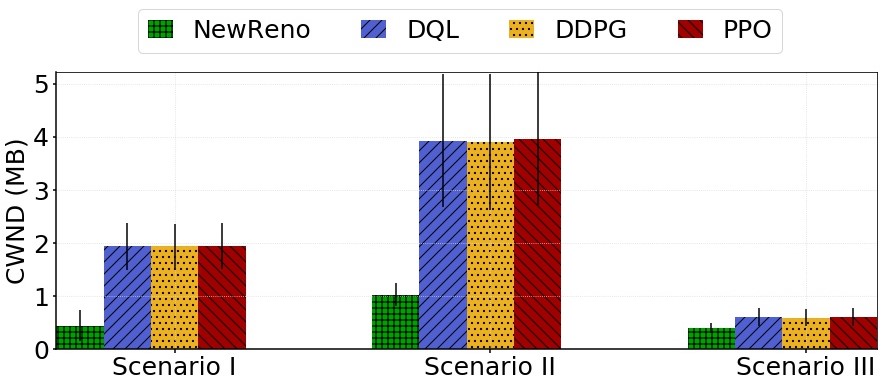}
			\vspace{-0.4cm}
			\caption{\textbf{CWND for the Three Scenarios}}
		\end{minipage}%
		\begin{minipage}[t]{0.5\linewidth}
			\vspace{-0.3cm}
			\centering
			\includegraphics[width=9cm,height=4cm]{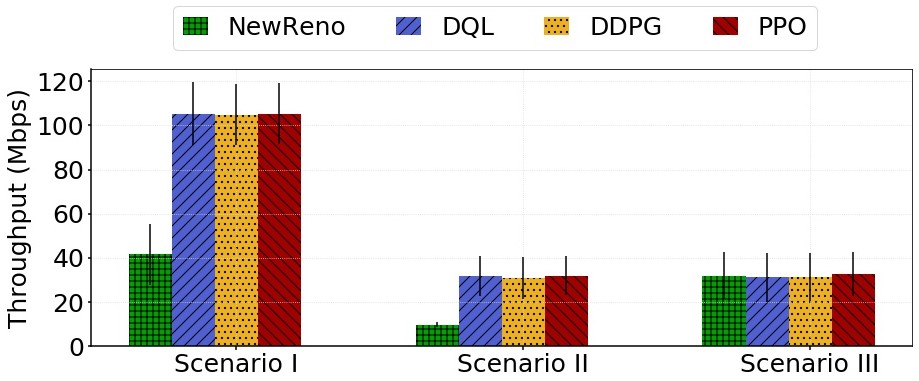}
			\vspace{-0.4cm}
			\caption{\textbf{Throughput for the Three Scenarios}}
		\end{minipage}%
		\centering
	\end{figure*} 
	
	\begin{figure*}[htbp]
		\begin{minipage}[t]{0.5\linewidth}
			\vspace{-0.3cm}
			\centering
			\includegraphics[width=9cm,height=4cm]{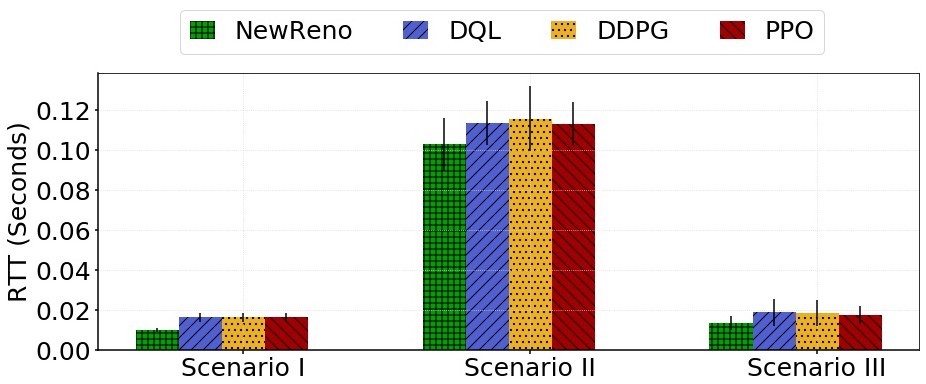}
			\vspace{-0.4cm}
			\caption{\textbf{RTT for the Three Scenarios}}
		\end{minipage}%
		\begin{minipage}[t]{0.5\linewidth}
			\vspace{-0.3cm}
			\centering
			\includegraphics[width=9cm,height=4cm]{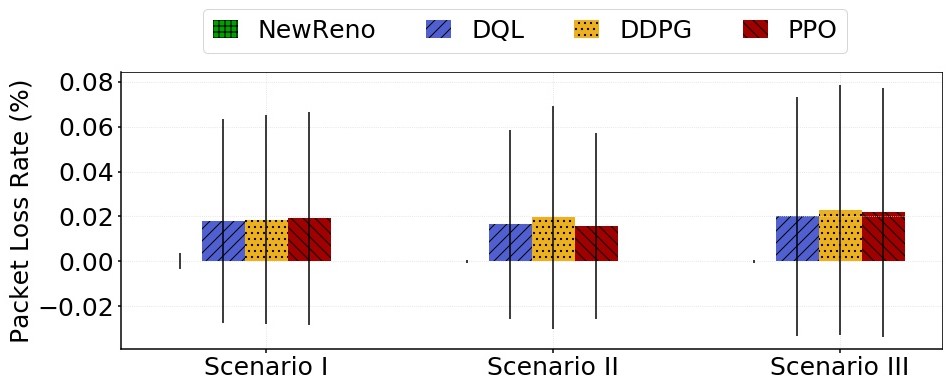}
			\vspace{-0.4cm}
			\caption{\textbf{Packet Loss Rate for the Three Scenarios}}
		\end{minipage}%
		\centering
	\end{figure*}

	\begin{figure}[]
		\vspace{-0.3cm}
		\centering
		\includegraphics[width=9cm,height=4cm]{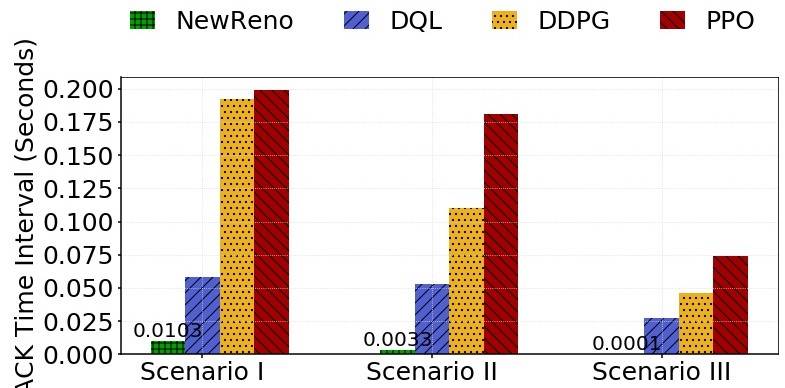}
		\vspace{-0.4cm}
		\caption{\textbf{ACK Interval for Realistic Network Simulation}}
	\end{figure}%
	
	\begin{figure*}[htbp]
		\begin{minipage}[t]{0.33\linewidth}
			\centering
			\includegraphics[width=6cm,height=3.3cm]{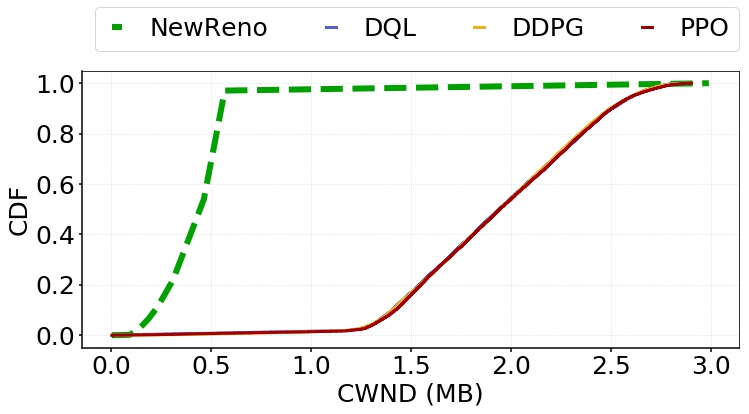}
			\caption{\textbf{CDF of CWND in Scenario I}}
			\label{}
		\end{minipage}
		\begin{minipage}[t]{0.33\linewidth}
			\centering
			\includegraphics[width=6cm,height=3.3cm]{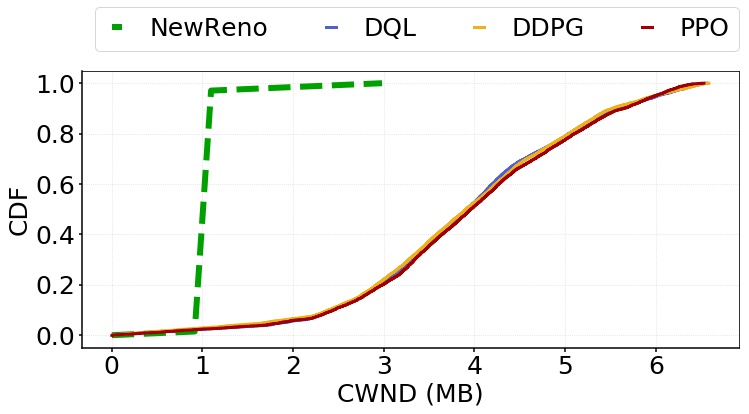}
			\caption{\textbf{CDF of CWND in Scenario II}}
		\end{minipage}%
		\begin{minipage}[t]{0.33\linewidth}
			\centering
			\includegraphics[width=6cm,height=3.3cm]{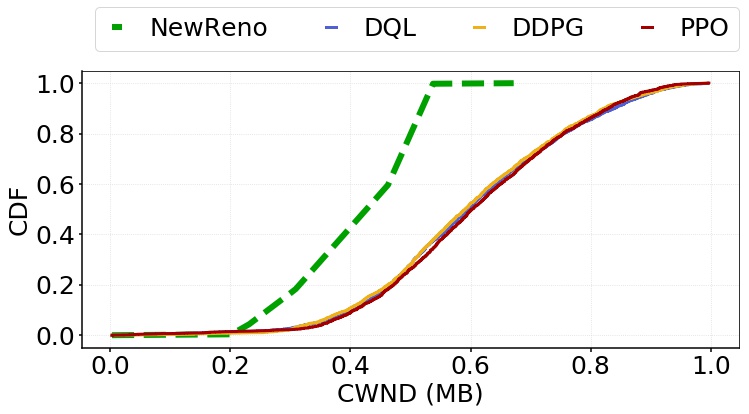}
			\caption{\textbf{CDF of CWND in Scenario III}}
		\end{minipage}%
		\centering
	\end{figure*} 
	
	\begin{figure*}[htbp]
		\begin{minipage}[t]{0.33\linewidth}
			\centering
			\includegraphics[width=6cm,height=3.3cm]{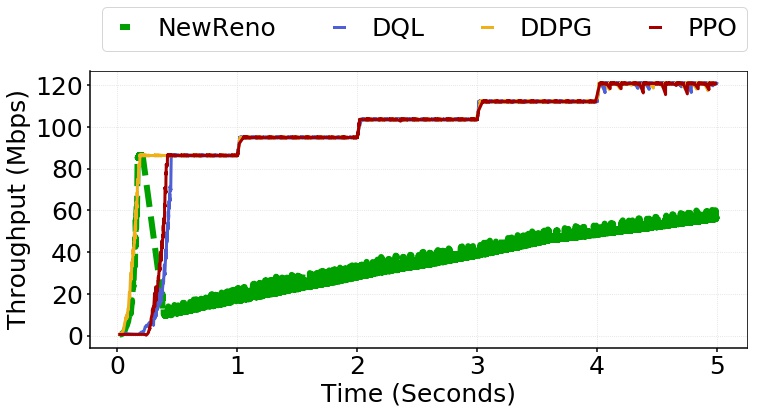}
			\caption{\textbf{Timeline of Throughput for Scenario I}}
			\label{}
		\end{minipage}
		\begin{minipage}[t]{0.33\linewidth}
			\centering
			\includegraphics[width=6cm,height=3.3cm]{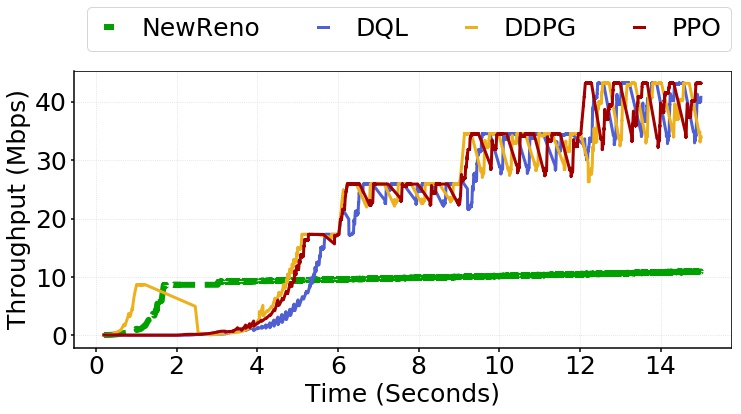}
			\caption{\textbf{Timeline of Throughput for Scenario II}}
		\end{minipage}%
		\begin{minipage}[t]{0.33\linewidth}
			\centering
			\includegraphics[width=6cm,height=3.3cm]{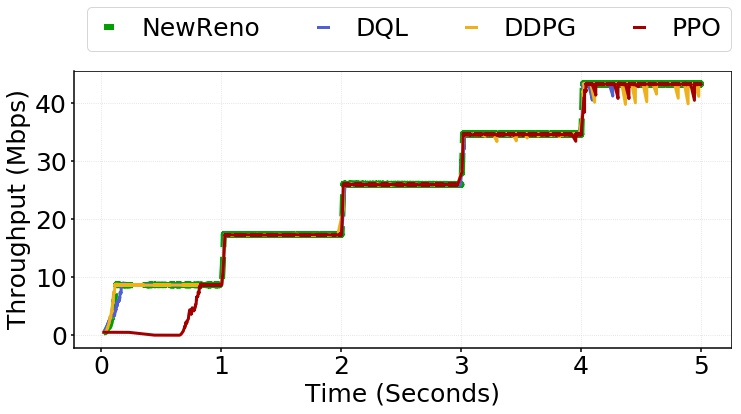}
			\caption{\textbf{Timeline of Throughput for Scenario III}}
		\end{minipage}%
		\centering
	\end{figure*} 
	
	\begin{figure*}[htbp]
		\begin{minipage}[t]{0.33\linewidth}
			\centering
			\includegraphics[width=6cm,height=3.3cm]{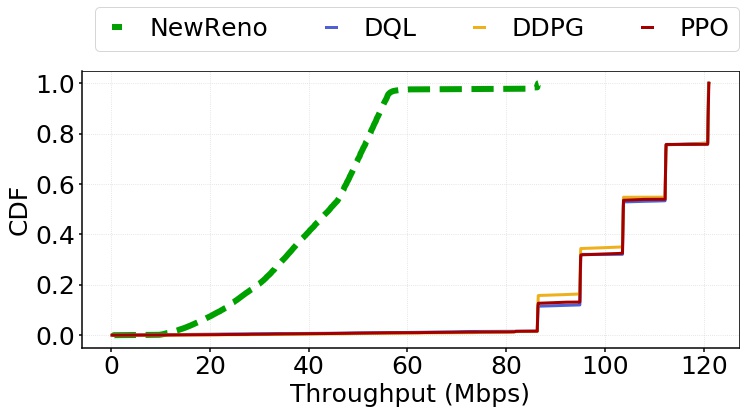}
			\caption{\textbf{CDF of Throughput for Scenario I}}
			\label{}
		\end{minipage}
		\begin{minipage}[t]{0.33\linewidth}
			\centering
			\includegraphics[width=6cm,height=3.3cm]{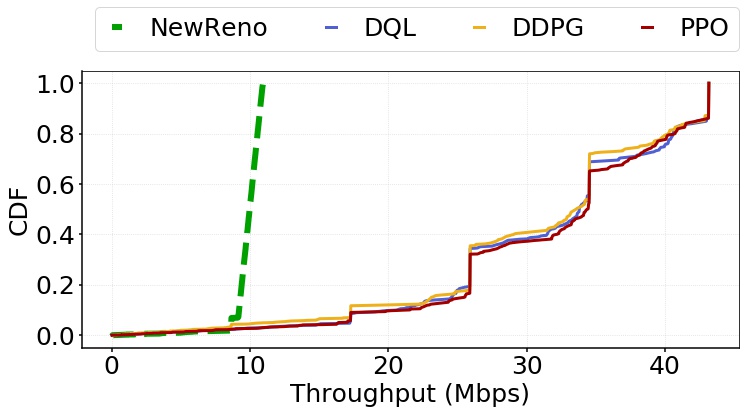}
			\caption{\textbf{CDF of Throughput for Scenario II}}
		\end{minipage}%
		\begin{minipage}[t]{0.33\linewidth}
			\centering
			\includegraphics[width=6cm,height=3.3cm]{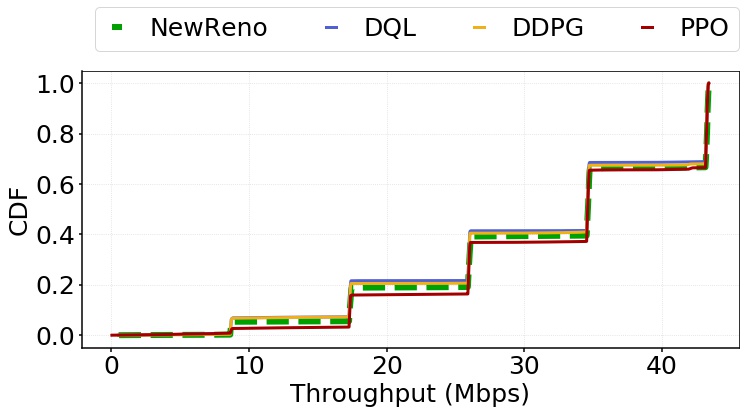}
			\caption{\textbf{CDF of Throughput for Scenario III}}
		\end{minipage}%
		\centering
	\end{figure*} 
	
	\begin{figure*}[htbp]
		\begin{minipage}[t]{0.33\linewidth}
			\centering
			\includegraphics[width=6cm,height=3.3cm]{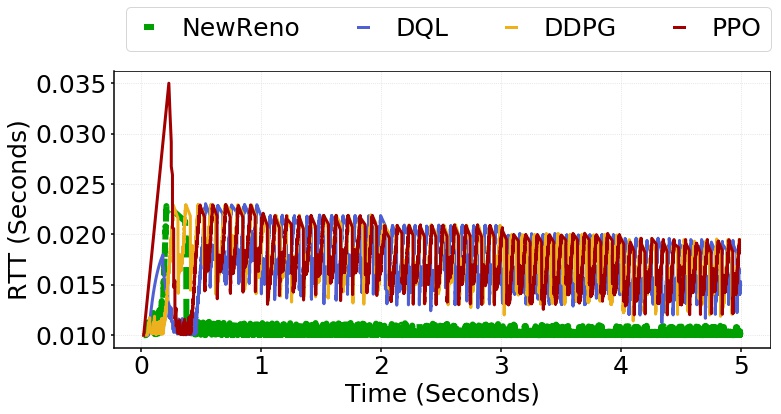}
			\caption{\textbf{Timeline of RTT for Scenario I}}
			\label{}
		\end{minipage}
		\begin{minipage}[t]{0.33\linewidth}
			\centering
			\includegraphics[width=6cm,height=3.3cm]{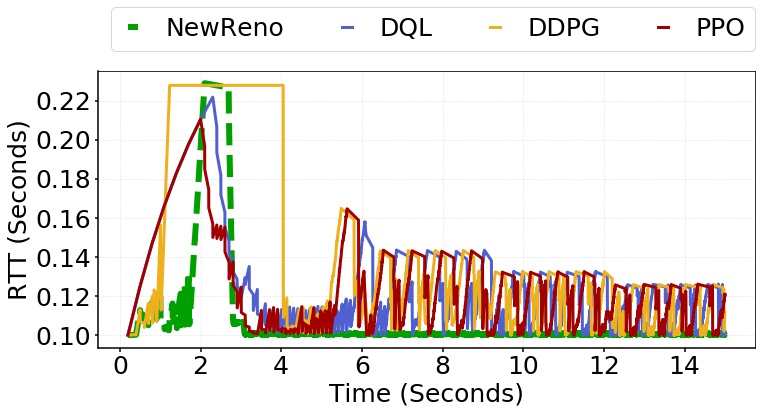}
			\caption{\textbf{Timeline of RTT for Scenario II}}
		\end{minipage}%
		\begin{minipage}[t]{0.33\linewidth}
			\centering
			\includegraphics[width=6cm,height=3.3cm]{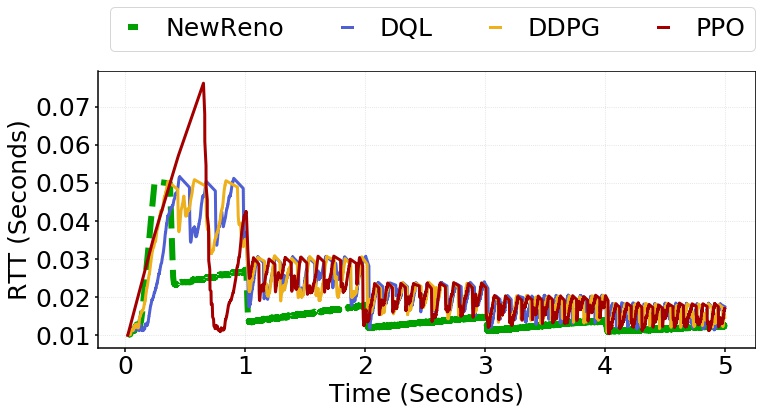}
			\caption{\textbf{Timeline of RTT for Scenario III}}
		\end{minipage}%
		\centering
	\end{figure*} 
	
	\begin{figure*}[htbp]
		\begin{minipage}[t]{0.33\linewidth}
			\centering
			\includegraphics[width=6cm,height=3.3cm]{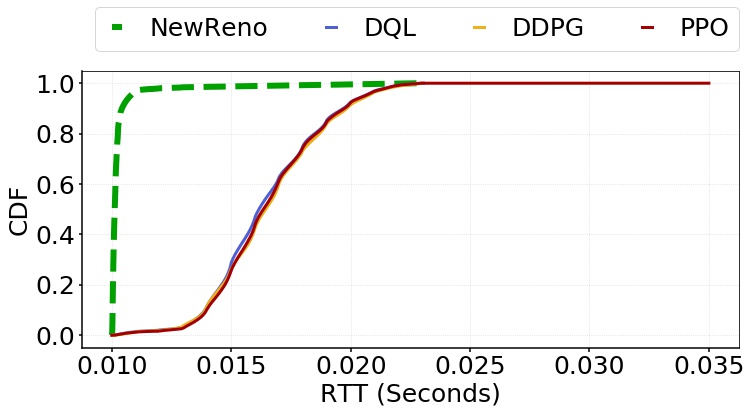}
			\caption{\textbf{CDF of RTT for Scenario I}}
			\label{}
		\end{minipage}
		\begin{minipage}[t]{0.33\linewidth}
			\centering
			\includegraphics[width=6cm,height=3.3cm]{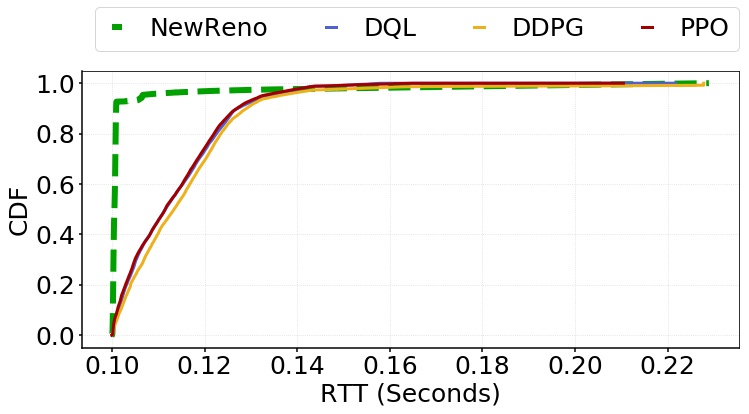}
			\caption{\textbf{CDF of RTT for Scenario II}}
		\end{minipage}%
		\begin{minipage}[t]{0.33\linewidth}
			\centering
			\includegraphics[width=6cm,height=3.3cm]{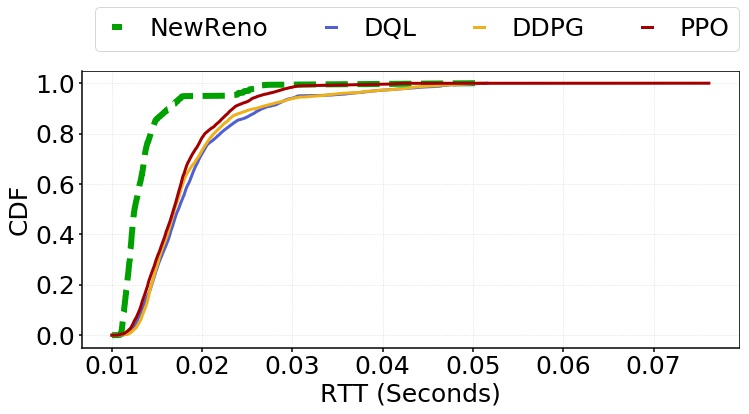}
			\caption{\textbf{CDF of RTT for Scenario III}}
		\end{minipage}%
		\centering
	\end{figure*} 
	
	\begin{figure*}[htbp]
		\begin{minipage}[t]{0.33\linewidth}
			\centering
			\includegraphics[width=6cm,height=3.3cm]{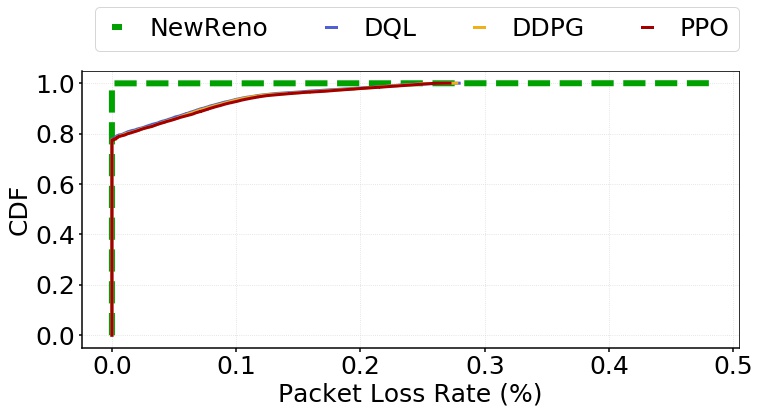}
			\caption{\textbf{CDF of Loss Rate for Scenario I}}
			\label{}
		\end{minipage}
		\begin{minipage}[t]{0.33\linewidth}
			\centering
			\includegraphics[width=6cm,height=3.3cm]{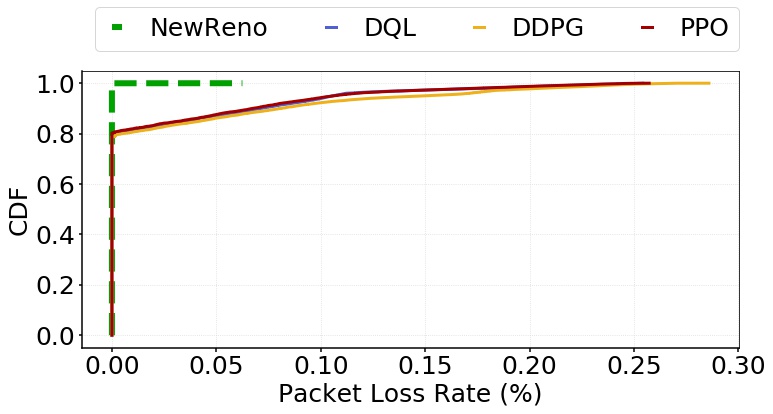}
			\caption{\textbf{CDF of Loss Rate for Scenario II}}
		\end{minipage}%
		\begin{minipage}[t]{0.33\linewidth}
			\centering
			\includegraphics[width=6cm,height=3.3cm]{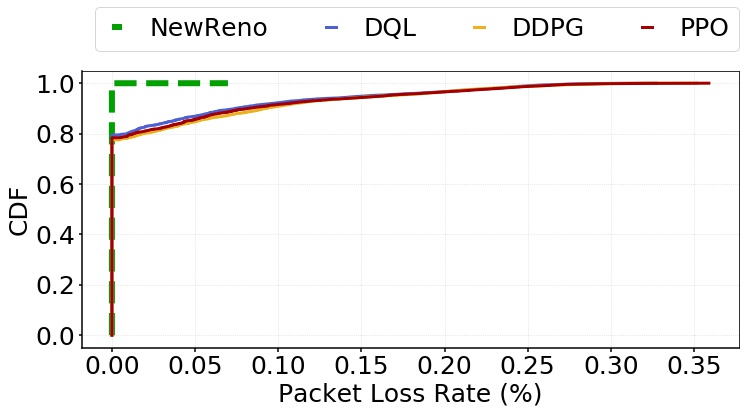}
			\caption{\textbf{CDF of Loss Rate for Scenario III}}
		\end{minipage}%
		\centering
	\end{figure*}

\subsection{Simulation Results}
The overall simulation results are shown in Table XI and Figures 8-30 including timeline figures showing the changes of performances, bar figures showing the average and the variances of performances and cumulative distribution function (CDF) figures showing the rough distributions of performances.

To check the performance of RL-based CC algorithms in realistic networks, we use Python to build sockets and send real data using the Linux platform. The result shows that the ACK interval is influenced by the computational complexity of the algorithms. As shown in Figure 12, the ACK interval of RL-based CC algorithms is much larger than NewReno, resulting in low increase in speed of CWND. Since RL-based CC algorithms require considerable time to calculate and obtain the action, the ACK will not be transferred in a timely fashion. It is noted that even amongst RL-based CC algorithms, there exist differences as well. Moreover, the delayed ACK influences the measurement of real throughput and RTT. Therefore, RL-based CC algorithms may be not applicable for realistic networks.

To compare performances of RL-based CC algorithms and NewReno, we show the output based on NS3 where the delay caused by the RL algorithms is excluded. In the following sections, the detailed performances are discussed including CWND and performance metrics.
	
\subsubsection{CWND}
Among the three RL-based CC algorithms, there is minimal difference between them in the three scenarios as shown in Figure 8. Moreover, we observe that the size of CWND of RL-based CC algorithms is much larger than rule-based CC algorithms in scenario I and scenario II, which both have high BPD as expected. While in scenario III, there is not much difference between these four algorithms. From Figure 13 to Figure 15, the CDF figures show distributions of CWNDs in three scenarios. As expected, in scenario I and scenario II, the sizes of CWND tend to be larger when RL-based CC algorithms are applied. 
	
\subsubsection{Throughput}
Theoretically, throughput of RL-based CC algorithms is expected to exceed the throughput of NewReno due to the increased average of CWND of RL-based CC algorithms in scenario I and scenario II. As shown in Figure 9, our speculation is verified. In scenario I and scenario II, throughput is  improved when the RL-based CC algorithms are used. While in scenario III, RL-based CC algorithms show no advantage. For the detailed distribution and timeline of throughput, more figures from Figure 16 to Figure 21 augment the results and explanation.  
	
\subsubsection{RTT}
The RTT of NewReno is small and stable, which represents the benchmark of RTT. In three scenarios, compared with NewReno, RTT is higher in networks with RL-based CC algorithms as shown in Figure 10,25,26 and 27. Because the increase of CWND is more aggressive among RL-based CC algorithms, it is understandable that RTT is higher. However, from the Figure 22 to 24, it shows that increments of RTT are limited and bounded compared with increments of throughput in scenario I and scenario II.
	
\subsubsection{Packet Loss Rate}
As shown in Figure 11, the packet loss rate of NewReno is almost zero while there are minimal packet loss in networks with RL-based CC algorithms. Moreover, the distribution information shows the increased packet loss rate in RL-based CC algorithms from Figure 28 to Figure 30. Considering the aggressiveness of RL-based CC algorithms, bounded packet losses are understandable.
	
\subsection{Analysis of Results}
	
From the simulation results, it can be seen that RL-based CC algorithms can achieve high throughput with limited increased RTT and packet loss rate in networks with relatively high BDP. Moreover, in our network environments, three RL-based CC algorithms exhibited similar performance. Because the space complexity was not so high and the dynamic fluctuation was limited, these three algorithms handled these scenarios well. Therefore, our experiments showed that RL-based CC algorithms have advantages in high BDP networks (as simulated using NS3).
	
In realistic network, CC algorithms react to the ACK arrival time. When a new ACK comes, the algorithm detects the delay or loss in the network and then adjusts the CWNDs or the sending rate. For traditional CC algorithms, there is a minor cost in time to compute the action because the adjustment rule are pre-designed and stable, while RL-based CC algorithms require lots of time to input the states to the neural network; get the output; update the action value and then take the appropriate actions. This process is clearly time-consuming especially with the potential size of ACK transmission rates. As such, it is hard for RL-based CC algorithms to measure the actual transmission time of ACKs and almost impossible to measure the real network throughput, RTT and packet loss rate. On the NS3 platform, these problems are not revealed because the NS3 platform  separates the computation  and  transmission parts. Therefore, no matter how time-consuming the algorithm is, there is no impact on the ACK transmission. However in real world applications, such time must be considered. Thus whilst RL-based CC algorithms are applicable on the NS3 platform, they are limited to realistic environments. 
	
\subsection{Proposed Solutions} 
	
Based on the simulation results and analysis, it can be observed that current RL-based CC algorithms process rewards based on arrival of ACKs, which are transferred and received one by one. As discussed, these RL-based CC algorithms are feasible on the NS3 simulator which separates the calculation and ACK transmission, however, the implementation of RL-based CC algorithms is still a problem. As a result, there are several possible  future research trends.

Firstly, designing lighter models based on mapping tables to deal with the problem of time-consuming RL-based CC algorithms. After a RL-based model is trained in network emulators, it can save the state and action to a table. Therefore, a mapping table can be prepared in advance. This process can be done off-line. When the model is deployed, only the mapping table is used. Given the state of the network environment, the action is given based on the mapping table. The time of this process is relatively small. This method can be efficient and time-saving. However, there are some challenges. The simple mapping table may be large and unwieldy in continuous scenarios. Therefore, more efficient mapping tables might be explored address these limitations.
	
Another solution is decreasing the frequency of decisions, such as employing RL to select CC algorithms in a given time interval instead of selecting CWND size based on ACK arrival intervals. This means that the time interval for updating is much larger than the delay caused by the calculation of RL. Therefore, the impact of the delay can be ignored. Of course, the drawback is that the updating speed and responsiveness of the RL algorithm would be slower. To balance these two performance issues, further research is required.
	
Finally, asynchronous RL algorithms are supposed to deal with delayed ACKs due to the algorithms' computational complexity. In an asynchronous RL framework, there are multiple actors. These actors take effect asynchronously, which can eliminate the effects of delayed ACKs. Therefore, in the network thread, ACKs are not blocked by the RL agent thread. In \cite{55}, to handle the delay of rewards, one action generates several partial actions. Therefore, each partial action can interact with the network environment independently. In addition, in \cite{129}, an asynchronous RL training framework, TorchBeast, combined with Pantheon network emulators, is used to handle delayed actions. The proposed algorithm, MVFST-RL, separates the network transmission and RL agents in realistic network communications based on multiple asynchronous actions. Though the algorithm eliminates the effect of delayed actions, the high resource-demanding training process is a problem since there are multiple actors to be trained and the state space is larger compared to synchronous RL-based CC algorithms. Therefore, the training process is more difficult. More research is required to address this issue.
	
\section{Challenges and Trends of Learning-based Congestion Control Schemes} 
	
\subsection{Challenges of Learning-based Congestion Control schemes}
For rule-based CC algorithms, the main issue is to detect congestion promptly and react quickly. The challenge of this kind of algorithm is dealing with flexibility. It is hard to satisfy different scenarios with a single algorithm. For learning-based CC algorithms, flexibility is improved but there are some issues that need to be addressed.
	
\textbf{Parameter Selection} influences the performance heavily especially with RL algorithms. State space, action space, reward design and other hyper-parameters related to algorithm structures need to be considered carefully. Using reward design as an example. In a RL-based CC algorithm, throughput and RTT are used to calculate the reward. While in other RL-based CC algorithms, packet loss rate and delay are considered when calculating the reward. For supervised learning, predefined parameters determine potential classification errors which affect the performance of CC. For unsupervised learning algorithms, parameters such as the number of clustering groups and initial cluster centers influence the final clustering results. Therefore, optimizing parameters is a non-trivial activity.
	
\textbf{High Computational Complexity} is a significant  issue for learning-based CC algorithms. For supervised learning techniques, especially for  hybrid and complex methods such as boosting and bagging, the prediction accuracy can be extremely high, but the computational complexity can also be high. For RL algorithms, the computational complexity results in delayed actions and rewards. This impacts on the utilization of  bandwidth. 
	
\textbf{High Memory Consumption} needs to be taken into consideration. The training of RL-based CC algorithms requires considerable storage space especially for continuous network environments. Therefore, abstracting the state-action space and obtaining representative data is needed for an efficient training process. For example, LSTM \cite{8} and Kanerva coding \cite{4} are used to represent and abstract the network states. Some advanced RL frameworks such as DDPG \cite{54} and A3C \cite{55} have a strong capability to deal with continuous network environments by representing the state-action space using complex neural networks.  Abstracting  representative state is thus key. Currently, a huge space representation is a major limitation of complex scenarios.
	
\textbf{Low Training Efficiency} is related to the feasibility of  deployment. For learning-based CC algorithms, the training process may be time- and resource-consuming. State abstraction is important to improve the training efficiency. Optimal parameter selection can be helpful to improve the training efficiency as well. Tackling this requires more research. Current learning-based CC algorithms require significant amounts of training data to guarantee the performance based on simulations. However, though diverse network topologies and traffic flows can be simulated, the algorithms cannot always avoid over- and under-fitting problems.
	
\textbf{Hard Convergence} impacts RL-based CC algorithms. Considering  complex algorithms with multiple neural networks, it can be difficult to attain convergence. Current RL algorithms propose different approaches to contribute to  convergence, however for realistic networking, this cannot always be guaranteed.

\textbf{Incompatibility} is an open question requiring future research. Current learning-based CC algorithms are often used as a built-in component or an independent controller to control congestion. There is still a long way to go for the issues related to compatibility between learning-based CC algorithms and traditional CC algorithms to be resolved.

\subsection{Trends of Learning-based Congestion Control Algorithms}
	
Considering the issues associated with learning-based CC algorithms as mentioned above, there are several trends that should be considered.
	
Firstly, engineering issues related with RL-based CC algorithms are a key research topic due to the high online capability of RL algorithms. Based on the previous literature, most RL-based CC algorithms are based on simulations using network emulators. On the one hand, simulations with network emulators eliminate unrelated factors and are more suitable to design network scenarios. On the other hand, engineering issues can be ignored, e.g. parameter selection and computational complexity. In realistic network communications, such engineering issues are significant for RL-based CC algorithms. To design more applicable algorithms, simulations in realistic network environments will be a primary focus moving forward. 
	
In addition, lightweight learning-based CC algorithms will be a hot topic in the future. Robust domain knowledge is needed to realize lightweight learning-based CC algorithms. Current learning-based CC algorithms have high complexity and can require considerable time to make decisions, with significant demands on memory and storage. Therefore, lighter-weight learning-based CC algorithms are required to be more applicable and deployable. To make models lighter, domain knowledge supporting  model-driven techniques look promising. Compared with the solid foundation of traditional CC algorithms which cover underlying theories such as RTT distributions in different scenarios and reordering schemes, current learning-based CC algorithms are relatively coarse-grained with limited knowledge support. Learning-based CC algorithms require a complete and detailed state space to train the model, making the model heavier. Lightweight models using fewer optimally chosen parameters is needed. 

Finally, an open network platform that provides massively differentiated dynamic network scenarios supporting the exploration and evaluation of various learning-based CC algorithms is needed to facilitate further research in learning-based CC algorithms. Pantheon \cite{144} belongs to this kind of platform. Though this platform covers diverse nodes, professional and specific network environments are not offered, e.g. flexible ad hoc wireless networks. Therefore, there is a demand for a general platform providing a professional and realistic simulation environment to train learning-based CC algorithms. In this way, the development of learning-based algorithms will be faster.
	
\section{Conclusion}
Due to the limitations of traditional CC algorithms in dynamic networks, learning-based CC algorithms have seen a recent trend in academia. In this paper, we provided a review of state of the art in learning-based CC algorithms together with simulations focused on different RL-based CC algorithms as representatives of learning-based CC algorithms are conducted. In the simulations, it was shown that RL-based CCs algorithms exhibit better performances compared to traditional CC algorithms in different scenarios such as networks with high bandwidth and low delay. We presented and discussed limitations with current RL-based CC algorithms for realistic deployments and outline some approaches that may be used in future research. We identified challenges and trends associated with learning-based CC algorithms including dealing with engineering issues related to RL-based CC algorithms. In the future, network environments are expected to be increasingly complicated. Given this, there is a clear need for addressing such complexity and flexibility. To improve the performance and robustness, further research is required to deal with issues such as computation time, data storage and pre-designed parameters. We argue that lightweight and efficient learning-based models with general learning-based platforms are needed and will be a future research focus.

	
	\bibliographystyle{IEEEtran}
	\bibliography{IEEEabrv,ref}
	
	

\end{document}